\documentclass[final,5p,times,twocolumn]{elsarticle}

\usepackage{ntheorem} % 替代方案

\usepackage[caption=false,font=footnotesize,labelfont=rm,textfont=rm]{subfig}

\usepackage{threeparttable}
\usepackage{color}
\usepackage{booktabs} % For formal tables
\usepackage{comment}
\usepackage{footmisc}
\usepackage{mathrsfs}
\usepackage{graphicx}
\usepackage{multirow}
\usepackage{algorithm}
\usepackage{multicol}  
\usepackage{enumitem}
\usepackage{array}
\usepackage{float}
\usepackage{textcomp}
\usepackage{stfloats}
\usepackage{url}
\usepackage{verbatim}
\usepackage{graphicx}
\usepackage{algpseudocode}

\usepackage{url}
\usepackage[pdfencoding=auto]{hyperref}

\usepackage{amsmath}   % 数学公式支持
\usepackage{amsfonts}  % 数学符号

\hypersetup{
    colorlinks=true,
    linkcolor=blue,
    filecolor=magenta,      
    urlcolor=cyan,
    citecolor=blue,
}

\journal{Accepted for publication in Expert Systems with Applications}

\begin{document}

\begin{frontmatter}

\title{FastPFRec: A Fast Personalized Federated Recommendation with Secure Sharing\tnoteref{arxiv_note}}
\tnotetext[arxiv_note]{This paper has been accepted for publication in Expert Systems with Applications.}

\author[label1,label2]{Zhenxing Yan}
\ead{23125335@bjtu.edu.cn}

\author[label1,label2]{Jidong Yuan\corref{cor1}} %% 正确放置 \corref
\ead{yuanjd@bjtu.edu.cn}

\author[label1,label2]{Yongqi Sun}
\ead{yqsun@bjtu.edu.cn}

\author[label1,label2]{Haiyang Liu}
\ead{haiyangliu@bjtu.edu.cn}

\author[label1,label2]{Zhihui Gao}
\ead{zhihuigao@bjtu.edu.cn}

\cortext[cor1]{Corresponding author: Jidong Yuan (yuanjd@bjtu.edu.cn)}

\affiliation[label1]{
  organization={Key Laboratory of Big Data \& Artificial Intelligence in Transportation (Beijing Jiaotong University)},
  addressline={Ministry of Education},
  city={Beijing},
  postcode={100044},
  % state={Beijing},
  country={China}
}

\affiliation[label2]{
  organization={School of Computer Science and Technology},
  addressline={Beijing Jiaotong University},
  city={Beijing},
  postcode={100044},
  % state={Beijing},
  country={China}
}

%% Abstract
\begin{abstract}
Graph neural network (GNN)-based federated recommendation systems effectively capture user--item relationships while preserving data privacy. However, existing methods often face {slow convergence} on graph data and {privacy leakage risks} during collaboration. To address these challenges, we propose \textbf{FastPFRec} ({\textbf{Fast} \textbf{P}ersonalized \textbf{F}ederated \textbf{Rec}ommendation with Secure Sharing}), a novel framework that enhances both {training efficiency} and {data security}. FastPFRec accelerates model convergence through an efficient local update strategy and introduces a privacy-aware parameter sharing mechanism to mitigate leakage risks. {Experiments on four real-world datasets (Yelp, Kindle, Gowalla-100k, and Gowalla-1m) show that} FastPFRec achieves {32.0\% fewer training rounds}, {34.1\% shorter training time}, and {8.1\% higher accuracy} compared with existing baselines. These results demonstrate that FastPFRec provides an efficient and privacy-preserving solution for scalable federated recommendation.
\end{abstract}


\end{frontmatter}

\section{Introduction}

Graph neural networks (GNNs) \cite{zhou2020graph} have emerged as a cutting-edge paradigm for modeling complex user-item interactions in recommendation systems, providing powerful capabilities for capturing graph-structured relationships. These models have been widely adopted in next-generation intelligent services, such as personalized content delivery across mobile and IoT platforms\cite{PARK2026129695}. Despite their remarkable effectiveness, GNNs predominantly rely on centralized access to user behavior data, which introduces critical privacy and security concerns in real-world deployments and highlights the urgent need for privacy-preserving learning frameworks that can maintain personalization without compromising sensitive information. This challenge is further exacerbated by stringent legal frameworks such as the GDPR \footnote{\url{https://gdpr-info.eu/}}, which mandates strict guidelines on the processing, storage, and sharing of personal data—particularly concerning user consent and data anonymization. Such regulations pose significant compliance challenges for data-driven industries that depend on large-scale user data to refine their services. In this context, federated learning (FL) \cite{li2020review} has arisen as a promising distributed training paradigm, enabling collaborative model training without exposing raw user data, thus preserving privacy and facilitating compliance with regulations. Notably, FL is increasingly being applied in consumer-centric domains—including personalized recommendations, smart health monitoring, and voice assistants to enhance service quality while systematically mitigating privacy risks.

\begin{figure}[t]
  \centering
  \includegraphics[width=0.49\textwidth]{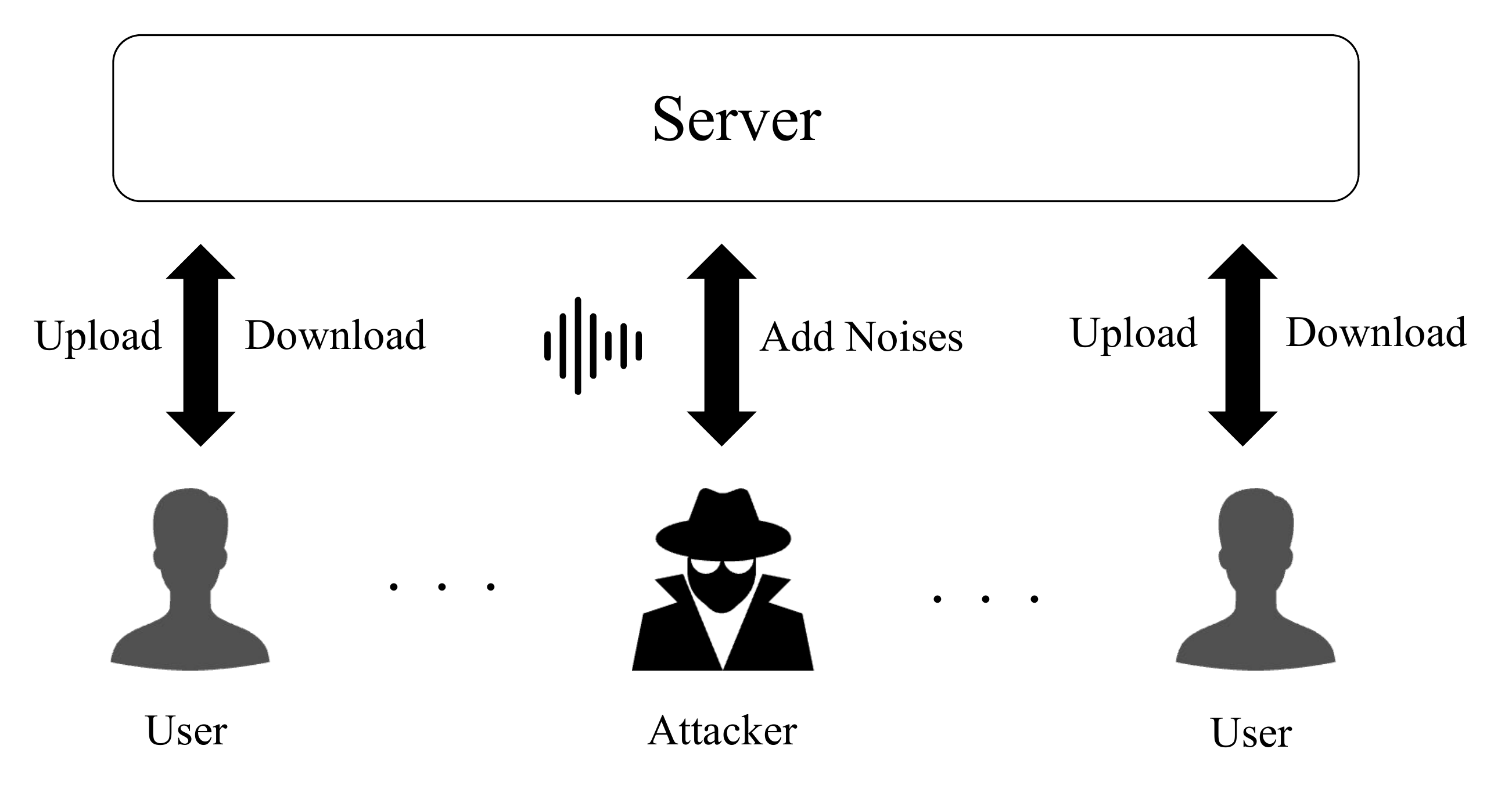}
  \caption{An example of federated attack with noise injection.}
  \label{attacker}
\end{figure}

However, integrating GNNs with FL also presents non-trivial challenges. First, the iterative process of neighborhood aggregation in GNNs, when performed in a federated setting with heterogeneous (Non-IID) user data, often leads to slow model convergence and substantial communication overhead \cite{luo2022hysage}. Second, while FL avoids raw data sharing, the model parameters exchanged during training could still potentially leak sensitive user information, necessitating stronger privacy guarantees beyond the basic FL architecture. For example, Figure~\ref{attacker} illustrates a scenario where an attacker directly targets the server by noisy injection, disrupting the entire federated recommendation system. If the server is compromised, the training process halts, and the model's integrity is at risk. Additionally, in cases where the server directly interacts with clients, it can trace the source of the uploaded parameters, potentially leading to privacy breaches. This highlights the need for robust security mechanisms to safeguard both system functionality and client privacy \cite{mothukuri2021survey}.

{
Standard two-tier FL architectures (client-server) lack robust mechanisms to detect and mitigate malicious or noisy client updates, which can degrade recommendation quality and compromise user privacy. Hierarchical FL methods \cite{liu2020client} reduce communication overhead through intermediate aggregation, but they focus primarily on efficiency rather than security, and do not address recommendation-specific privacy risks such as inference attacks on user-item interactions. Existing federated recommenders (e.g., PerFedRec\cite{luo2022personalized}, FedGNN\cite{wu2021fedgnn}) rely on standard FL architectures without specialized security mechanisms for GNN-based recommendation scenarios. This work addresses these gaps by proposing a three-tier federated architecture with trusted nodes that combine the communication efficiency of hierarchical FL with security-focused mechanisms tailored for recommendation systems.
}

\begin{figure}[h]
  \centering
  \includegraphics[width=0.45\textwidth]{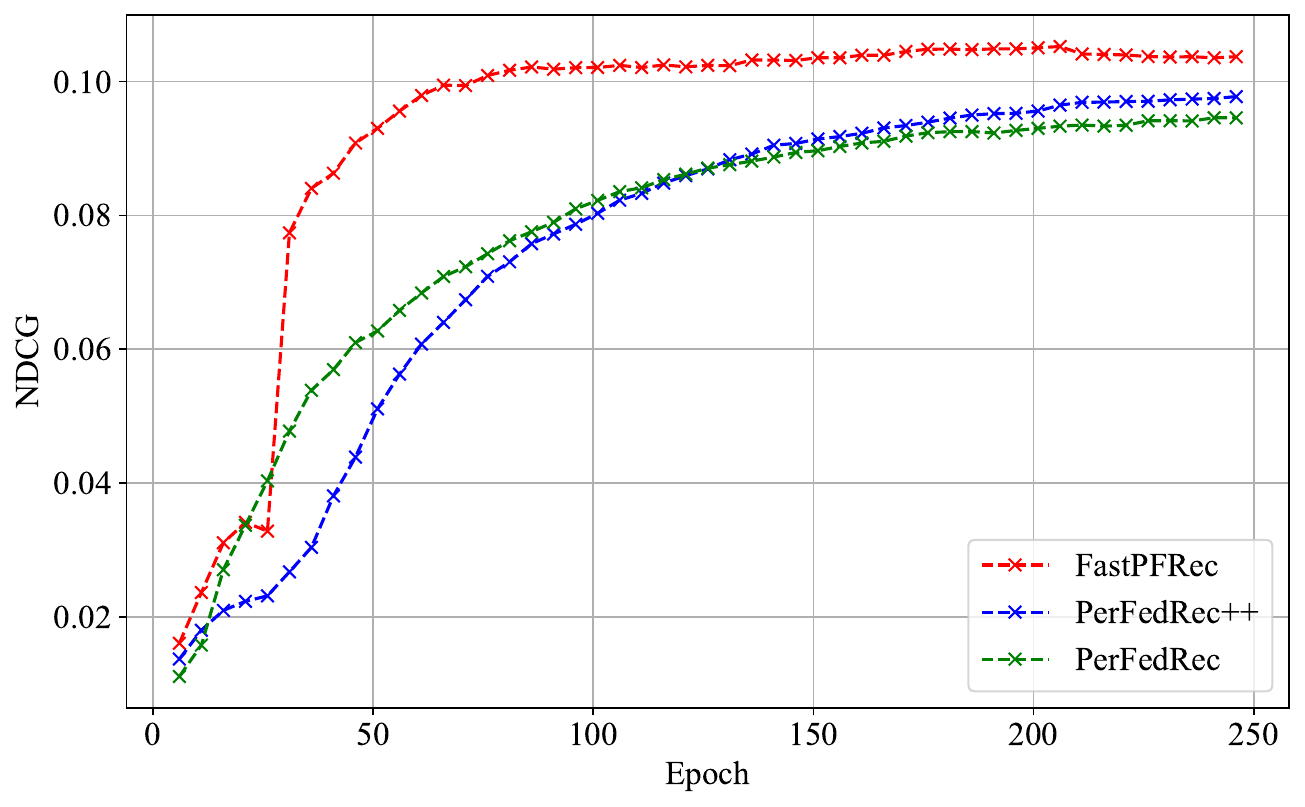}
  % \vspace{-3mm}
  \caption{Comparison of training NDCG across epochs on the Gowalla-1m dataset.}
  \label{PerFedRec_plus_ndcg}
  % \vspace{2mm}
\end{figure}

Beyond privacy and security concerns, recent studies in federated recommendation systems have primarily focused on enhancing model performance and reducing communication costs \cite{banabilah2022federated, lin2020meta}. Several approaches, such as federated averaging \cite{li2020secure, mcmahan2017communication} and secure aggregation protocols \cite{fereidooni2021safelearn}, have been proposed to accelerate convergence and ensure privacy protection. Although these methods have achieved promising metric results, the convergence issue remains unresolved. Figure~\ref{PerFedRec_plus_ndcg} illustrates the relationship between the number of training epochs and the evaluation metric, normalized discounted cumulative gain (NDCG), for the state-of-the-art (SOTA) personalized federated recommendation methods, PerFedRec \cite{luo2022personalized} (green line) and PerFedRec++ \cite{luo2024perfedrec++} (blue line), on the large-scale Gowalla-1m dataset \cite{liang2016modeling}. As shown, both PerFedRec and PerFedRec++ exhibit slow convergence, with their performance not reaching optimal levels even after nearly 300 epochs, by which point the training time has already exceeded 5 hours. After analysis, it is clear that this is because the local GNN model is complex and the local user vectors have not been sufficiently trained.

To address these issues, we propose a novel federated recommendation method called \textbf{FastPFRec} (\textbf{\underline{Fast}} \textbf{\underline{P}}ersonalized \textbf{\underline{F}}ederated \textbf{\underline{Rec}}ommendation with secure sharing), which is designed to accelerate the convergence speed of federated GNNs while ensuring the privacy of user data. A key component of {FastPFRec} is the proposed FastGNN, a simplified and fast-converging GNN update scheme that significantly improves the training efficiency. By integrating FastGNN, the number of training epochs required is greatly reduced, leading to a lower time complexity. As shown in Figure~\ref{PerFedRec_plus_ndcg}, {FastPFRec} (red line) achieves superior performance within just 100 epochs, consistently outperforming other methods in terms of NDCG at every epoch.  Furthermore, we introduce a new parameter upload mechanism that ensures data privacy within FL frameworks, effectively preventing user data leakage.
Specifically, inspired by hierarchical FL \cite{liu2020client}, we introduce the concept of {trusted nodes} within the FL system. These trusted nodes act as intermediaries between clients and the server, securely aggregating the model parameters uploaded by the clients. Our main contributions are summarized as follows:

\begin{itemize}[leftmargin=*, itemsep=0pt, topsep=0pt]

  \item  We propose {FastPFRec}, a three-tier federated recommendation framework with {trusted nodes} that perform secure intermediate aggregation and anomaly detection, going beyond standard two-tier FL recommenders.
  \item  We design a {FastGNN update schedule} that updates user embeddings frequently while refreshing item embeddings sparsely, cutting effective training cost without sacrificing accuracy—distinct from existing GNN-based FL methods that fully update both sides each round.
  \item We combine {graph perturbation}, {local differential privacy (LDP)}  \cite{truex2020ldp} , and {trusted-node aggregation} to strengthen robustness against noisy clients, with an explicit threat model.
  \item Extensive experiments on real-world datasets show significantly faster convergence and improved recommendation quality compared with SOTA GNN-based federated recommenders.

\end{itemize}

\section{Related Work}
This section reviews the literature on two key areas: GNN-based federated recommendation systems and the privacy-preserving techniques employed within this domain.

\subsection{GNN-Based Federated Recommendations}
GNN-based federated recommendation systems learn from decentralized user-item interaction graphs. In this paradigm, a central server aggregates local model updates, and privacy is often enhanced using techniques like differential privacy or homomorphic encryption to protect sensitive data \cite{li2022federated, li2020federated}. Many such systems employ graph convolutional networks (GCNs) to effectively model complex user-item relationships, enabling secure and personalized recommendations \cite{zhang2019graph}. For example, {FedGNN} \cite{wu2021fedgnn} pioneered the use of GNNs to model user-item feature interactions in a federated context. Building on this, {FeSog} \cite{liu2022federated} integrated social network information to capture inter-user dynamics. For enhanced personalization, {FedGCLRec} \cite{SAMAD2026129294} dynamic social influence learning with personalized federated optimization, while {GPFedRec} \cite{zhang2024gpfedrec} constructs a user-relation graph from personalized item embeddings to preserve data locality. These methods use GCN and its variants, and require a relatively high number of training rounds to reach convergence.

Other approaches focus on user clustering. {PerFedRec} \cite{luo2022personalized} utilizes $K$-means clustering \cite{ahmed2020k} to group users by behavioral similarity, enabling group-specific recommendations. Its successor, {PerFedRec++} \cite{luo2024perfedrec++}, improves upon this by incorporating contrastive learning in a pre-training phase, which refines the distinction between user and item representations and boosts performance. 

Despite their effectiveness, these models often exhibit growing structural complexity and prolonged training times, leading to significant computational overhead. To mitigate these challenges of slow convergence and model complexity \cite{zhang2020batchcrypt}, we propose {FastGNN} which simplifies the GCN update method.

\subsection{Privacy Protection in Federated Recommendation}
Privacy is a cornerstone of federated recommendation\cite{KIM2025128568}. By design, raw user data remains on local devices, with only model updates (e.g., gradients or parameters) being transmitted to the server \cite{yang2020federated, ma2020safeguarding}. To further secure these updates during transmission, cryptographic methods like differential privacy \cite{dwork2006differential} are commonly applied to obfuscate the gradients.

Early work such as FCF \cite{ammad2019federated} focused on processing item gradients to prevent leakage of raw ratings, while FedMF \cite{chai2020secure} protects privacy by encrypting item embeddings during transmission. FedRec \cite{lin2020fedrec} enhanced privacy by mitigating data reconstruction risks through a hybrid strategy for handling missing ratings and sampling non-interacted items. Subsequent techniques include FedNCF \cite{jiang2022fedncf}, which applies differential privacy to computed gradients to defend against inference attacks. Most recently, Fedai \cite{HE2025126564} leverages homomorphic encryption-based grouping and interaction anonymization to balance recommendation accuracy with robust privacy preservation.

To further secure computations, previous works \cite{feng2018privacy} have proposed privacy-preserving tensor decomposition over encrypted data in federated clouds and secure two-party inference frameworks like Panther \cite{feng2025panther}. While these cryptographic methods provide strong security guarantees, they often incur significant computational and communication overheads compared to lightweight perturbation methods. Consequently, recent advancements have focused on refining LDP mechanisms to better balance data utility and privacy. For instance, Zhang et al. \cite{zhang2025privacy} proposed a mixture model-based matrix factorization framework under LDP, effectively addressing user heterogeneity. Similarly, in location-based scenarios, Zhang et al. \cite{zhang2025maximizing} introduced a prior knowledge-enhanced geo-indistinguishable approach to demonstrate the versatility of differential privacy in complex spatial tasks. Parallel to confidentiality, robustness against manipulation is also critical. Recent works \cite{zhao2025attack, fang2025dmia} have explored attack-agnostic defense frameworks to protect LDP-based systems against manipulation attacks.

However, a significant vulnerability persists in these models: the central server can often trace parameter updates back to specific clients. Additionally, malicious clients can directly attack the server, exposing users to potential inference attacks \cite{yang2021toward}. To address this critical gap, we introduce a novel three-tier architecture comprising {clients}, {trusted nodes}, and a {server}.

{HierFAVG \cite{liu2020client} use intermediate edge servers to reduce communication overhead. While our trusted-node mechanism shares the hierarchical structure, it differs in focus: hierarchical FL optimizes for communication efficiency, whereas our trusted nodes prioritize security and privacy protection through anomaly detection and secure aggregation, specifically designed for GNN-based recommendation scenarios.}

In our framework, clients send perturbed model updates to trusted nodes, which aggregate these updates before forwarding them to the server. This intermediate aggregation layer effectively decouples clients from the server, preventing source tracing and enhancing user anonymity, scalability, and efficiency.

\begin{figure*}[t]
\centering
\includegraphics[width=1.0\linewidth]{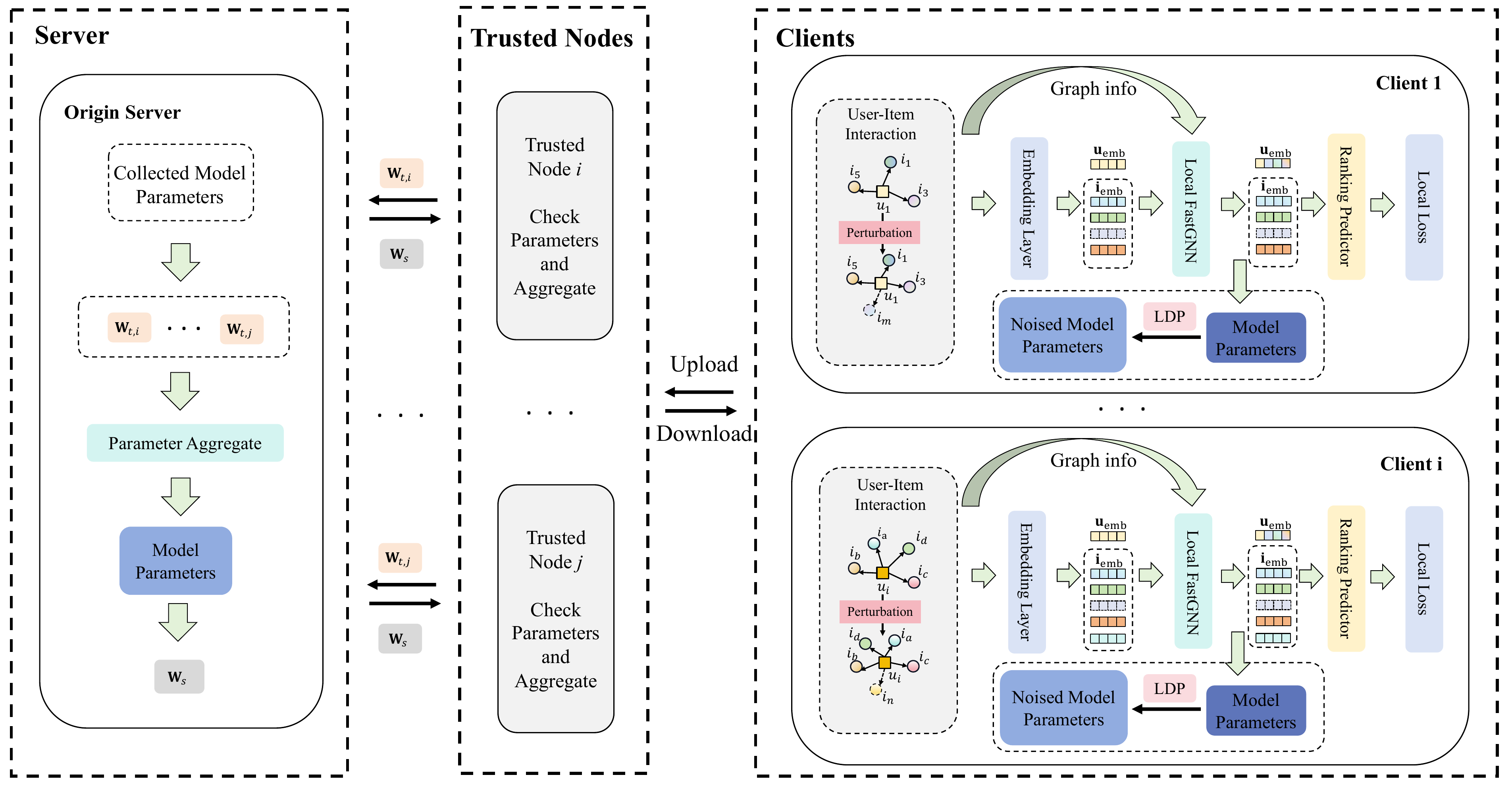}
\caption{The proposed \textbf{FastPFRec} framework. This flowchart illustrates the end-to-end process, highlighting the local training with \textbf{FastGNN} on client devices and the subsequent aggregation of model updates by \textbf{trusted nodes} before being sent to the central server.}
\label{fig_workflow}
\end{figure*}

\section{Method}

\begin{algorithm}[H]
  \caption{FastPFRec}
  \label{algorithm:FastPFRec}
  \begin{algorithmic}[1]
    \Require Local graph $G_{u,i}$, random threshold $\tau$, noise scale $\lambda$, update weight $\beta$, max rounds $Q$, learning rate $\alpha$, distance threshold $\mu$, outlier ratio $\nu$
    
    \Ensure Local model parameters $\mathbf{\Theta}_i$ for all clients
    
    \State \textbf{Step 1: Client-Side Training}
    \ForAll{client $c_i \in \{1,2,\dots,C\}$ \textbf{in parallel}}
      \State Build local graph  and do perturbation
      \State Train local FastGNN and update embeddings: $\mathbf{u}_{\mathrm{emb}}, \mathbf{i}_{\mathrm{emb}} \gets \text{GNN}(G'_{u,i})$
      \State Add noise: $\mathbf{\Theta}'_i \gets \mathbf{\Theta}_i + \eta, \eta \sim \text{Lap}(\lambda)$
      \State Upload $\mathbf{\Theta}'_i$ to trusted node 
    \EndFor
    
    \State \textbf{Step 2: Trusted Node Aggregation}
    \For{$i = 1$ to $T$}
      \State Receive $C_i$ clients' parameters $\mathbf{\Theta}'$
      \If{$\neg \text{Check}(\mathbf{\Theta}', \mathbf{W}_{s}, \mu, \nu)$}
        \State $\mathbf{W}_{t,i} \gets \frac{1}{C_i} \sum_{j=1}^{C_i} \mathbf{\Theta}'_j$
        \State Send $\mathbf{W}_{t,i}$ to server
      \EndIf
    \EndFor
    
    \State \textbf{Step 3: Server-Side Global Update}
    \State $\mathbf{W}_s \gets \frac{1}{T} \sum_{j=1}^{T} \mathbf{W}_{t,j}$
    \State Distribute $\mathbf{W}_s$ to clients
    
    \State \textbf{Step 4: Client Model Update}
    \ForAll{client $c_i$ \textbf{in parallel}}
      \State $\mathbf{\Theta}_i^{q+1} \gets (1-\beta) \mathbf{\Theta}_i^q + \beta \mathbf{W}_s$
    \EndFor
    
    \State \textbf{End.}
  \end{algorithmic}
\end{algorithm}

This section provides a detailed overview of the FastPFRec framework, covering client-side, trusted node, and the server-side operational methods. The overall process is presented in Figure \ref{fig_workflow} and the algorithm procedure is shown in Algorithm~\ref{algorithm:FastPFRec}.

\subsection{Problem Definition}\label{section_definition}

In our personalized federated recommendation, we consider multiple clients, each representing a user and its interaction records. The goal is to train a federated recommendation system while ensuring data privacy. Let $U = \{u_1, u_2, \dots, u_N\}$ and $I = \{i_1, i_2, \dots, i_M\}$ represent the user and item sets, respectively. The user-item interactions form a bipartite graph $G_{u,i}$, where $y_{ui} = 1$ indicates an interaction between user $u$ and item $i$. We define the mapping $f_{\text{pred}}(\cdot)$ as a function that predicts the rating $\hat{y}_{ui}$ for a given user $u$ and item $i$ and $F$ as a loss function.

\subsection{Client-Side Method}
The client-side approach aims to achieve localized training of the user-item interaction graph while preserving user data privacy. This method primarily includes mechanisms for local data perturbation, the generation of embedding representations and the design of FastGNN. The detailed client-side training process is outlined in lines 1-7 of Algorithm~\ref{algorithm:FastPFRec} and illustrated in the right half of Figure~\ref{fig:trusted_node_mine}. The following sections provide a detailed description of the implementation of each step.

\subsubsection{User-Item Graph Construction as Input}

On the client side \(c_i\), each user has their local user-item interaction data, which can be represented as a user-item interaction graph \(G_{u,i}\). The formal representation of this graph is
\begin{equation}
G_{u,i} = (\mathcal{V},\mathcal{E}),
\end{equation}
where \(\mathcal{V} = \{u, i_1, i_2, \dots, i_n\}\) is the set of nodes in the graph, including the user \(u\) and the set of items $\mathcal{N}(u) = \{i_1, i_2, \dots, i_n\}$ with which the user interacts; \(\mathcal{E} = \{e_{u,i_j} \mid i_j \in \mathcal{N}(u) \}\) is the set of edges, representing the interaction between user \(u\) and item \(i_j\); Each edge \(e_{u,i_j} \in \mathcal{E} \) is associated with a weight \(w_{u,i_j}\), which could represent a binary interaction (clicked/not clicked) or a rating value. For each user \(u\), the client locally stores \(G_{u,i}\) as the input for subsequent model training.

\subsubsection{Client-side Privacy Protection Mechanism}

We randomly add perturbations to some users by treating items that the user has not interacted with as items the user has interacted with. This effectively obscures the user's true behavior, thereby enhancing data privacy.

The perturbation process modifies the graph by randomly adding false interactions. Specifically, for a given item \( i_j \in I \), if the user \( u \) has not interacted with \( i_j \) (i.e., \( e_{u,i_j} \notin E_u \)), we randomly decide with a certain probability \( P_{\text{pert}} \) whether to add the edge \( e_{u,i_j} \). This means that some items the user has not interacted with will be treated as items the user has interacted with, thus effectively perturbing the original graph. Mathematically, the perturbation process can be described as

\begin{equation}
G_{u,i}' = G_{u,i} \cup \{ e_{u,i_j} \mid i_j \in I, e_{u,i_j} \notin G_{u,i}, \text{and} \, P_{\text{pert}}(i_j) > \tau \},
\end{equation}
where \( E_u' \) is the perturbed interaction set, \( P_{\text{pert}}(i_j) \) is the perturbation probability for item \( i_j \), \( \tau \) is a random number uniformly sampled from the interval [0, 1]. 
{
In our experimental setup, we set the perturbation probability parameter $P_{\text{pert}} = 0.1$ to quantify the intensity of the perturbation. This implies that approximately 10\% of the non-interacting items are randomly selected and added as false edges to the user's local graph $G_{u,i}$.
}
In this process, if \( P_\text{pert}(i_j) > \tau \), the edge \( e_{u,i_j} \) will be added to the graph, even if the user has not previously interacted with item \( i_j \). And we set $\mathcal{N}'(u)$ a union of previously user $u$ interacting items and disturbed items. This perturbation mechanism ensures that the original user-item interaction graph is modified, preventing the accurate reconstruction of the user's true interaction history, thereby enhancing data privacy \cite{lin2020fedrec}.

\subsubsection{Message Passing in Local FastGNN}

The embedding representations of users and items serve as the foundation for subsequent GNN training. Locally, the client maps nodes (users and items) to a low-dimensional vector space via an embedding layer. For a user \( u \) and an item \( i_j \), their embedding representations are

\begin{equation}
\mathbf{u}_{\mathrm{emb}} = f_{\mathrm{emb}}(u), \quad \mathbf{i}_{j, \mathrm{emb}} = f_{\mathrm{emb}}(i_j),
\end{equation}
where \( f_{\text{emb}} \) is the embedding generation function; The dimension of the embedding vectors is \( k \), which is typically a hyperparameter. The embedding vectors \( \mathbf{u}_{\mathrm{emb}} \) and \( \mathbf{i}_{j, \mathrm{emb}} \) are generated through random initialization or pre-training methods and serve as the initial inputs to the GNN.

In the local training phase, the client uses a simplified FastGNN to model the user-item graph \( G_{u,i} \) and capture the higher-order interactions between users and items. Instead of aggregating the user and item vectors at every update, we perform convolutional aggregation on the data at intervals. {Specificially, the item embeddings are updated only every $ H \times 10 $ epochs, while user embeddings are updated at every step. This design is based on the empirical observation that user representations require more frequent updates to capture dynamic preferences, while item representations are relatively stable. The interval of $   H \times 10 $ was chosen to strike a balance between efficiency and model performance: a shorter interval increases computational cost with diminishing returns, while a longer interval may lead to outdated item representations that harm accuracy.} This interval-based aggregation method not only reduces the computational overhead of each update but also improves training efficiency while maintaining model performance, providing a more practical solution for applications on large-scale datasets.

The message passing mechanism is used to update the embedding representations of users and items. At the \( l+1 \)-th layer, the update rule for the embedding representation of the user node \( u \) is

\begin{equation}
\mathbf{u}_{\mathrm{emb}}^{(l+1)} = \sigma \left( \sum_{i_j \in \mathcal{N}'(u)} \alpha_{u,i_j} \cdot \mathbf{i}_{j, \mathrm{emb}}^{(l)} \right),
\end{equation}
where $\mathcal{N}'(u)$  is the set of neighboring items for user $u$,   $\sigma$ is the nonlinear activation function (e.g., the \( \text{sigmoid} \) function).

After $H$ layers of message passing, the final embedding representations of the user and item are

\begin{equation}
\mathbf{u}_{\text{emb}} = \mathbf{u}_{\text{emb}}^{(H)}, \quad \mathbf{i}_{j,\text{emb}} = \mathbf{i}_{j,\text{emb}}^{(H)},
\end{equation}
these embedding representations are used for subsequent ranking predictions.

The accelerated convergence of FastGNN is achieved through a scheduled item-update mechanism, which strategically reduces computational overhead. This approach is grounded in the observation that item embeddings, being shared across multiple users, typically exhibit greater stability and evolve more slowly than user-specific embeddings during training. By updating item embeddings only every $H \times h$ epochs while continuing to update user embeddings at every epoch, FastGNN significantly lowers the average computational cost per epoch without compromising essential gradient information.

To formalize this efficiency gain, we analyze the computational complexity relative to standard GNN-based federated recommenders (e.g., PerFedRec++). Under standard assumptions including a fixed embedding dimension $k$, stable local graph sizes ($|U|$ users and $|I|$ items per client), and the typical bipartite interaction structure---a full GNN propagation step has a complexity of $\mathcal{O}(H \cdot |U| \cdot |I| \cdot k)$ for $H$ layers. In conventional methods, this cost is incurred in every epoch over $E$ local training rounds, leading to a per-round complexity of $\mathcal{O}(E \cdot H \cdot |U| \cdot |I| \cdot k)$.

In contrast, FastGNN's scheduled update strategy modifies this cost structure. Since user embeddings are updated every epoch but item embeddings are updated only approximately $E / (H \times h)$ times, the per-round complexity becomes:
\begin{equation}
\mathcal{O}\left(E \cdot H \cdot |U| \cdot k + \frac{E}{H \times h} \cdot H \cdot |I| \cdot k\right) = \mathcal{O}\left(E \cdot H \cdot k \cdot \left(|U| + \frac{|I|}{h}\right)\right).
\end{equation}
For common recommendation scenarios where $|I| \gg |U|$, this translates to an effective average cost reduction for item-related operations from $\mathcal{O}(H \cdot |U| \cdot |I| \cdot k)$ to roughly $\mathcal{O}(H \cdot |U| \cdot |I| \cdot k / h)$. It is important to clarify that this represents a reduction in the effective average computational cost per epoch, not an alteration of the asymptotic complexity order of a single update operation.

In this formulation, $h$ serves as a critical hyperparameter governing the trade-off between computational efficiency and the freshness of item representations. A larger $h$ further reduces costs but risks staleness, while a smaller $h$ approaches the cost of full updates. Based on our empirical ablation studies, we adopted $h=10$ as the optimal setting, as it effectively maximizes training speed improvements while maintaining sufficient embedding accuracy for robust recommendation performance.

Ranking prediction aims to predict the user's preference score for an item and rank the items accordingly. Specifically, the predicted preference score for user \( u \) and item \( i_j \) is calculated as

\begin{equation}
\hat{y}_{u,i_j} = f_{\text{pred}}(\mathbf{u}_{\text{emb}}, \mathbf{i}_{j,\text{emb}}),
\end{equation}
where \( f_{\text{pred}}(\cdot, \cdot) \) is the scoring function, typically defined as the dot product

\begin{equation}
f_{\text{pred}}(\mathbf{u}_{\text{emb}}, \mathbf{i}_{j,\text{emb}}) = \mathbf{u}_{\text{emb}}^\top \cdot \mathbf{i}_{j,\text{emb}}.
\end{equation}

To optimize the model parameters, we employ the bayesian personalized ranking (BPR) loss function, which is designed to learn user preferences by maximizing the probability that a user prefers a positive item over a negative item. The BPR loss is defined as:

\begin{equation}
\mathcal{L}_{\text{BPR}} = -\log(\sigma(\hat{y}_{ui} - \hat{y}_{uj})) + \gamma \cdot \frac{1}{N} \sum_{i=1}^{N} ||\mathbf{\Theta}_i||_2,
\end{equation}
where \( \hat{y}_{ui} \) and \( \hat{y}_{uj} \) represent the predicted scores for positive and negative items respectively, $j$ is a randomly sampled negative item that user $u$ has not interacted with, \( \sigma(\cdot) \) is the sigmoid function, \( \gamma \) is the regularization parameter, and \( ||\mathbf{\Theta}_i||_2 \) is the L2 norm of the model parameters.

Based on the predicted scores \( \hat{y}_{u,i_j} \), the items \( \{i_1, i_2, \ldots, i_n\} \) are sorted in descending order of the scores to generate a recommendation list.

\subsubsection{Model Parameter Upload and Download}

After training, the client uploads the model parameters to a trusted node. The specific process is as follows.

First, to protect the user’s local data privacy, the client applies LDP \cite{truex2020ldp} mechanism before processing or uploading local
data. Laplace noise is added to the model parameters \( \mathbf{\Theta}_c \):

\begin{equation}
\mathbf{\Theta}_c' = \mathbf{\Theta}_c + \mathbf{\eta},
\end{equation}
where \( \mathbf{\eta} \sim Lap(\lambda) \).

Then, the trusted node $i$ receives the parameters uploaded by all relevant clients and aggregates them to generate the global parameters.

The client constructs a local user-item interaction graph, applies the LDP mechanism for data security, uses embedding generation and FastGNN to capture interaction relationships, and generates personalized recommendations. Finally, the client uploads the perturbed model parameters for aggregation and downloads the global parameters to update the local model, balancing privacy protection and recommendation performance.

\subsection{Trusted Node}

\begin{figure}[t] % 
\centerline{\includegraphics[width=0.45\textwidth]{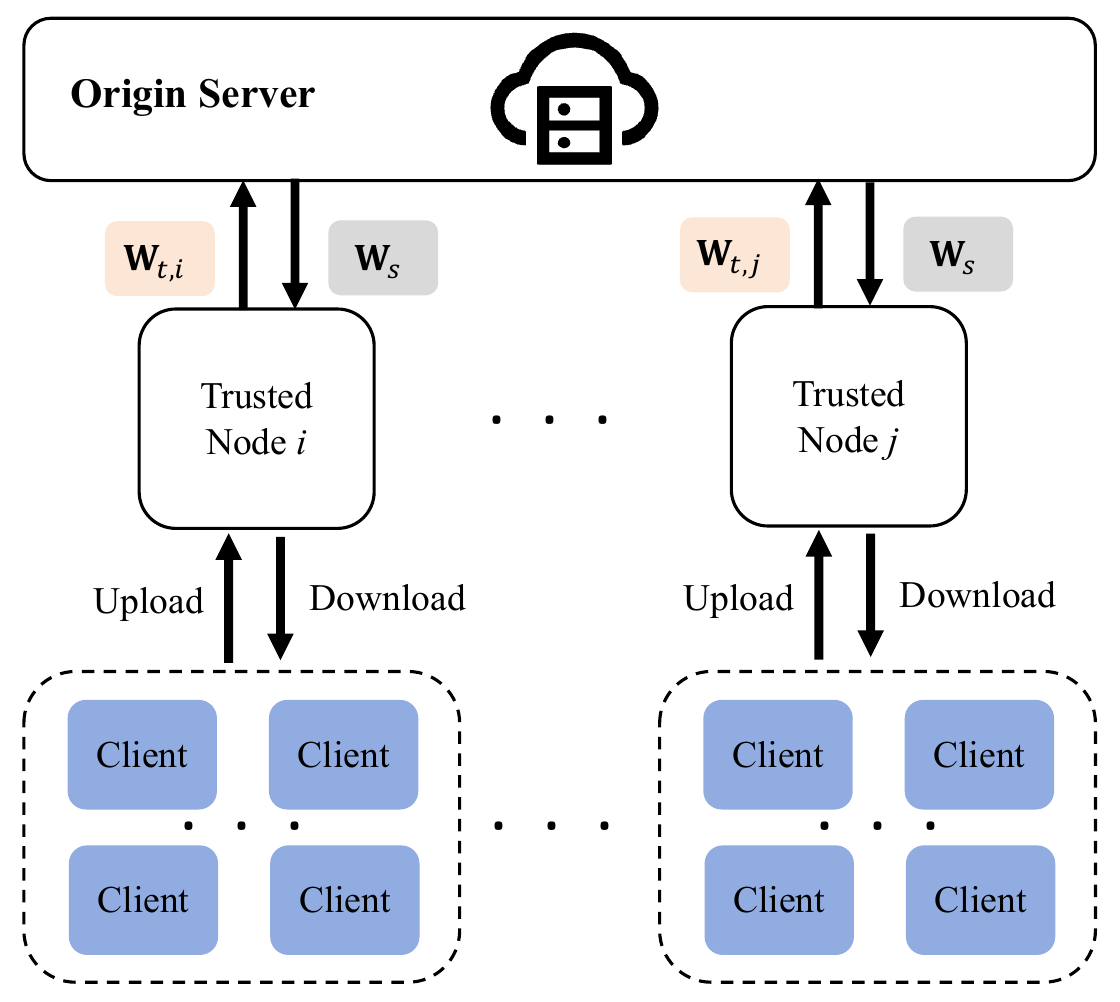}}
  % \vspace{-3mm}

\caption{The position of trusted node in the overall framework.}

\label{fig:trusted_node_mine}
  % \vspace{2mm}

\end{figure}

{In this three-tier architecture federated recommendation system, the trusted node serves as an intermediary between the client and the server, responsible for aggregating model parameters and ensuring privacy protection. By offloading the parameter aggregation process, it reduces the computational burden on the server, while also enhancing the overall security and robustness of the system. The subsequent sections detail the design, data flow, and privacy protection mechanisms employed by the trusted node, with further specifics provided in lines 8-15 of Algorithm~\ref{algorithm:FastPFRec} and Figure~\ref{fig:trusted_node_mine}.}

\subsubsection{Receiving Disturbed Parameters from Clients}

After client \( c_i \) completes local training, it randomly uploads the disturbed model parameters \( \mathbf{\Theta}_i' \) to the trusted node \( i \). The number of clients handled by trusted node \(i\) is denoted as $C_i$, and the total number of clients handled by all trusted nodes satisfies \(C_1 + C_2 + \ldots + C_T = C\). Each trusted node receives a set of disturbed parameters from multiple clients: $\{\mathbf{\Theta}_1', \mathbf{\Theta}_2', \dots, \mathbf{\Theta}_{C_{i}}'\}$, where \( C_i \) is the clients uploading to trusted node $i$.

% we employ a simplified version of this method\cite{nardi2022anomaly}
Before aggregation, each trusted node performs anomaly detection on the received parameters: 

\begin{equation}
\text{Check}(\mathbf{\Theta}', \mathbf{W}_{s}, \mu, \nu) = \frac{1}{C_i}\sum_{j=1}^{C_i} \mathbf{1}[|z_j| > \mu] > \nu,
\end{equation}
where $z_j$ is the robust Z-score for client $j$, $\mu$ is the distance threshold, and $\nu$ is the outlier ratio threshold. If malicious behavior is detected, the node is isolated and excluded from the global update process. To reduce the amount of data uploaded to the original server, the trusted node performs an initial aggregation of the disturbed parameters from the \( C \) clients it receives. The aggregation process can use a simple averaging operation

\begin{equation}
\mathbf{W}_{t,i} = \frac{1}{C_i} \sum_{j=1}^{C_i} \mathbf{\Theta}_j',
\end{equation}
where \( \mathbf{W}_{t,i} \) represents the result of the aggregation after the trusted node \( i \); \( C_i \) is the number of clients in the current round. The aggregated \( \mathbf{W}_{t,i} \) is then uploaded to the original server for global aggregation.

\subsubsection{Uploading and Distributing}

The trusted node uploads its aggregated parameters \( \mathbf{W}_{t,i} \) to the original server as one of the inputs for global parameter updates. Suppose there are \( T \) trusted nodes in the system, then the set of aggregated parameters received by the original server is $\{ W_{t,1}, W_{t,2}, \dots, W_{t,T} \}$.

As indicated in lines 17-20 of Algorithm ~\ref{algorithm:FastPFRec}, after the trusted node receives the global model parameters \( \mathbf{W}_s \) from the original server, it distributes them to the set of responsible clients for the local model updates. Each client updates its local model parameters as

\begin{equation}
\mathbf{\Theta}_i^{q+1} = (1 - \beta) \mathbf{\Theta}_i^q + \beta \mathbf{W}_s,
\end{equation}
where $ \mathbf{\Theta}_i^{q+1}$ is the local model parameter of client $ c_i$ after the update in round $ q+1$; $ \mathbf{\Theta}_i^q$ is the local model parameter of client $ c_i$ in round $q$; $ \beta \in (0,1)$ is  a hyperparameter, determining the weight of the global parameters in updating the local model.

\subsubsection{Privacy Protection Mechanism of Trusted Node}

The introduction of trusted nodes establishes a multi-layered privacy and security framework that substantially strengthens system robustness. {Regarding deployment conditions,} Trusted Nodes can be implemented as secure edge servers managed by the service provider. Unlike volatile client devices, these infrastructure-level entities possess stable power supplies and sufficient computational resources to execute robust aggregation and anomaly detection protocols, acting as reliable bridges between user terminals and the central cloud. {Under our trust assumptions, these nodes operate in an ``honest-but-curious'' threat model; they faithfully execute protocols but may be curious about client data.} Client-uploaded parameters, already perturbed via the LDP mechanism, are aggregated at trusted nodes, which perform statistical anomaly detection using median absolute deviation (MAD) to identify and isolate malicious updates. This process prevents compromised parameters from propagating to the central server. Notably, the architecture ensures that neither the server nor any single trusted node has direct access to clients' raw data; only perturbed, aggregated updates are processed, {ensuring strict privacy even against curious nodes}.

Beyond privacy, the three-tier design inherently mitigates single-point-of-failure risks {and robustly handles various failure scenarios}. With clients uniformly distributed across multiple trusted nodes, each node handles approximately $|C|/T$ clients. In the event of node failures, the system continues operating with the remaining nodes, affecting only the clients assigned to the failed nodes. Our experiments confirm that with $T=10$ trusted nodes, the system tolerates up to 3 node failures (30\% failure rate) without significant performance degradation. Even when trusted nodes themselves are compromised, the impact remains contained: a compromised node can observe only its assigned client subset, and its influence is diluted through server-level aggregation across all nodes. Client-side LDP provides an additional layer of protection, preserving privacy even if up to 30\% of nodes are compromised. To further bolster resilience, the system integrates automatic node isolation and dynamic client reassignment, ensuring continuous and secure operation under adversarial conditions.

\subsection{Server-Side Method}
The design goal of the server-side method is to achieve global model collaborative optimization by aggregating model parameters from multiple clients while ensuring both model accuracy and privacy protection. The server consists of the origin server and multiple trusted nodes, which work together to process the perturbed parameters uploaded by the clients and generate the optimized global model parameters. The following provides a detailed explanation of the processes involved in data upload, parameter aggregation, and global updates. The relevant pseudocode is provided by lines 16-22 of Algorithm~\ref{algorithm:FastPFRec}.

\subsubsection{Parameter Aggregation}
The trusted node is responsible for receiving the perturbed model parameters \( \mathbf{\Theta}_i' \) from multiple clients and uploading them to the origin server for global parameter aggregation. Let there be \( T \) trusted nodes, and the perturbed parameters uploaded by them are denoted as \( W_{t,1}, W_{t,2}, \dots, W_{t,T} \). The origin server aggregates all the parameters using the following equation:

\begin{equation}
\mathbf{W}_s = \frac{1}{T} \sum_{i=1}^{T} \mathbf{W}_{t,i},
\end{equation}
where \( \mathbf{W}_s \) is the aggregated global model parameters; ${W}_{t, i}$ is the perturbed parameters from the \( i \)-th trusted node.
This process uses averaging to integrate the local training results from all trusted nodes while effectively smoothing the impact of the noise, thus generating a global model $\mathbf{W}_s$, which is then provided to the clients as an update parameter.

\subsubsection{Global Parameter Distribute}
After the parameter aggregation, the origin server sends the global model parameters \( \mathbf{W}_s \) back to the trusted node, which then distributes them to the respective clients. And the trusted node will update the client parameters.
By aggregating parameters on a central server, the model can learn from diverse client data without exposing sensitive information. The trusted node acts as an intermediary, ensuring that the global model parameters $ \mathbf{W}_s $ are securely distributed to clients. This decentralized approach allows for the continuous updating of client models, ensuring that they stay up-to-date with global improvements while preserving privacy and reducing data transfer overhead. Thus, it enhances model performance while maintaining data confidentiality.

\subsection{Convergence Analysis Framework}
\label{subsec:convergence}

This section establishes the theoretical convergence guarantee for the FastPFRec algorithm under non-convex settings, accounting for Non-IID, LDP noise, and graph perturbations.

\subsubsection{Assumptions}
\label{subsubsec:assumptions}

Our analysis relies on the following standard assumptions in federated learning theory \cite{li2020federated, li2019convergence}:

\begin{enumerate}[
    leftmargin=*, 
    topsep=0pt, 
    itemsep=2pt, 
    partopsep=0pt, 
    parsep=0pt, 
    label=(\textbf{A\arabic*}), 
    series=assumptions
]
    \item \textbf{${L}$-smoothness:} 
    Each local loss function $F_i$ is $L$-smooth, i.e., there exists a constant $L > 0$ such that for any parameters $\mathbf{W}_1, \mathbf{W}_2$:
    \begin{equation}
        \begin{split}
            F_i(\mathbf{W}_1) \leq F_i(\mathbf{W}_2) 
            & + \langle \nabla F_i(\mathbf{W}_2), \mathbf{W}_1 - \mathbf{W}_2 \rangle \\
            & + \frac{L}{2} \|\mathbf{W}_1 - \mathbf{W}_2\|^2.
        \end{split}
    \end{equation}
    This ensures the loss function has Lipschitz-continuous gradients.

    \item \textbf{Bounded Gradient Variance:} 
    The stochastic gradient $g_i(\mathbf{W})$ satisfies:
    \begin{equation}
        \begin{gathered}
            \mathbb{E}[g_i(\mathbf{W})] = \nabla F_i(\mathbf{W}), \\
            \mathbb{E}[\|g_i(\mathbf{W}) - \nabla F_i(\mathbf{W})\|^2] \leq \zeta^2,
        \end{gathered}
    \end{equation}
    where the first condition ensures unbiased estimation, and the second bounds the variance.

    \item \textbf{Bounded Gradient Dissimilarity:} 
    For any $\mathbf{W}$, the divergence between local and global gradients is bounded:
    \begin{equation}
        \frac{1}{C} \sum_{i=1}^{C} \|\nabla F_i(\mathbf{W})\|^2 
        \leq \Gamma \|\nabla F(\mathbf{W})\|^2 + \delta^2,
    \end{equation}
    where $F(\mathbf{W}) = \frac{1}{C}\sum_{i=1}^C F_i(\mathbf{W})$ is the global loss function. 
    This characterizes the data heterogeneity across clients.

    \item \textbf{Bounded LDP Noise:} 
    The Laplace noise $\eta_i \sim \text{Lap}(\lambda)$ added for LDP has zero mean and bounded variance:
    \begin{equation}
        \mathbb{E}[\|\eta_i\|^2] = 2d\lambda^2 = \sigma_\eta^2,
    \end{equation}
    where $d$ is the model dimension and $\lambda$ controls the privacy level.

\item \textbf{Bounded Perturbation Impact:} 
The graph perturbation causes bounded parameter deviation:
\begin{equation}
    \mathbb{E}[\|\mathbf{\Theta}_i'^q - \mathbf{\Theta}_i^q\|^2] \leq \sigma_p^2,
\end{equation}
where $\mathbf{\Theta}_i'^q$ and $\mathbf{\Theta}_i^q$ are parameters from perturbed and original graphs, respectively.
\end{enumerate}

\subsubsection{Derivation of the Convergence Bound}
\label{subsubsec:proof}

The convergence proof establishes an upper bound for the expected squared gradient norm. Beginning with the global model update rule after $E$ local steps:

\begin{equation}
\mathbf{W}_s^{q+1} = \frac{1}{T} \sum_{i=1}^{T} \mathbf{W}_{t,i}^q
\quad\text{where}\quad 
\mathbf{W}_{t,i}^q = \frac{1}{C_i} \sum_{j=1}^{C_i} (\mathbf{\Theta}_j^q + \eta_j^q),
\end{equation}
we apply the $L$-smoothness property to analyze the loss improvement per communication round:
\begin{equation}
\begin{split}
\mathbb{E}[F(\mathbf{W}_s^{q+1})] \leq \mathbb{E}[F(\mathbf{W}_s^q)] 
    & - \alpha E \mathbb{E}[\|\nabla F(\mathbf{W}_s^q)\|^2] \\
    & + \frac{L}{2} \mathbb{E}[\|\mathbf{W}_s^{q+1} - \mathbf{W}_s^q\|^2].
\end{split}
\end{equation}
The key step involves bounding the variance term:
\begin{equation}
\begin{split}
\mathbb{E}[\|\mathbf{W}_s^{q+1} - \mathbf{W}_s^q\|^2] 
    &= \mathbb{E}\left[\left\| \frac{1}{T} \sum_{i=1}^{T} \left( \frac{1}{C_i} \sum_{j=1}^{C_i} (\mathbf{\Theta}_j^q - \mathbf{W}_s^q + \eta_j^q) \right) \right\|^2\right] \\
    &\leq G^2 + \sigma_\eta^2,
\end{split}
\end{equation}
where $G^2$ bounds the local update deviation and $\sigma_\eta^2$ is the LDP noise variance.

By recursively applying this inequality over $Q$ rounds, we obtain the final convergence bound:
\begin{equation}
\begin{split}
\frac{1}{Q} \sum_{q=0}^{Q-1} \mathbb{E}[\|\nabla F(\mathbf{W}_s^q)\|^2] 
    &\leq \frac{2(F(\mathbf{W}_s^0) - F^*)}{\alpha E Q} \\
    &\quad + C_1 \zeta^2 + C_2 \delta^2 + C_3 (\sigma_\eta^2 + \sigma_p^2).
\end{split}
\end{equation}

This result demonstrates that FastPFRec converges to a neighborhood of a stationary point, with the $\mathcal{O}(1/Q)$ term showing convergence rate and the remaining terms forming an error floor determined by system variances.

\section{Experiments}
In this section, we first outline the datasets and experimental details. Furthermore, we present and discuss the robustness and convergence effectiveness of FastPFRec across various baselines.

\begin{table}[h]
\centering
\caption{Statistics of datasets. Note that Gowalla-100k and Gowalla-1m are {dense subsets} derived from the original Gowalla dataset to simulate different scales.}
\label{dataset}
\resizebox{\columnwidth}{!}{

{
\begin{tabular}{c|cccc}
\toprule
Dataset & \# User & \# Item & \# Interaction & Density \\ \midrule
Yelp & 5,145 & 7,421 & 124,036 & 0.32\% \\
Kindle & 7,650 & 9,173 & 137,124 & 0.20\% \\
Gowalla-100k (Subset) & 3,917 & 7,000 & 85,442 & 0.31\% \\
Gowalla-1m (Subset) & 29,858 & 40,981 & 810,128 & 0.04\% \\
\bottomrule
\end{tabular}
}
}
\end{table}

\subsection{Experimental Settings}\label{section_exp_set}
\subsubsection{Datasets and Evaluation Metrics}

{To assess the performance of the proposed framework, we carried out experiments on four real-world datasets: Yelp\footnote{https://www.yelp.com/dataset}, Kindle\footnote{https://jmcauley.ucsd.edu/data/amazon/}, Gowalla-100k and Gowalla-1m\footnote{https://snap.stanford.edu/data/loc-gowalla.html}. The Gowalla-100k and Gowalla-1m datasets are particularly important for verifying the model's convergence on large-scale sparse datasets in real-world scenarios.} They are representative of data typically generated by consumer electronics devices, with their statistics summarized in Table \ref{dataset}. Yelp \cite{yelpdataset}  comes from the Yelp platform and contains multiple types of user-generated data. Kindle \cite{singh2024amazon} comes from Amazon's review data, covering user purchase records, ratings, and other information. Gowalla \cite{ference2013location} comes from the social sign from the platform Gowalla, which records the user's sign-in data of the geographical location in the real world. To verify the model's convergence on large-scale sparse datasets in real-world scenarios, we selected the Gowalla-100k and Gowalla-1m datasets. The density of Gowalla-1m is 0.0368\%.

{Regarding the nature of user-item interactions, Yelp and Kindle provide explicit feedback (ratings/reviews), while Gowalla-1m offers implicit feedback (location check-ins). We selected these datasets for two critical reasons. First, the underlying content is highly sensitive, making robust privacy protection a practical necessity. Second, their extreme sparsity closely simulates real-world federated environments, where client data is inherently localized and fragmented. Evaluating FastPFRec under these challenging conditions rigorously demonstrates its ability to extract robust global patterns from isolated, privacy-constrained data.}

{Gowalla-100k and Gowalla-1m:} Derived from the Gowalla check-in dataset (Feb. 2009--Oct. 2010), these are randomly sampled subsets with interaction scales approximating 100k and 1m, respectively. {Gowalla-100k} serves as a medium-scale, relatively dense benchmark (0.31\% density), while {Gowalla-1m} represents a large-scale, highly sparse scenario (0.04\% density). This distinction allows us to evaluate FastPFRec's robustness across varying data scales and sparsity levels. Detailed statistics are listed in Table \ref{dataset}.

In the evaluation, we employed two standard metrics including hit rate (HR) and NDCG, to evaluate the performance of top-10 recommendations \cite{wang2013theoretical}.

\subsubsection{Baselines}

\begin{table*}[t]
\centering
\caption{Performance comparison of different models on Yelp, Kindle, and Gowalla datasets (including Gowalla-100k and Gowalla-1m). Results are reported as mean $\pm$ standard deviation across 5 runs. ``*'' indicates statistical significance at $p < 0.05$ in Wilcoxon signed-rank test against baseline methods.}
\resizebox{\textwidth}{!}{
\begin{tabular}{cccccccccc}
\toprule
\multirow{2}{*}{Model} & \multicolumn{2}{c}{Yelp} & \multicolumn{2}{c}{Kindle} & \multicolumn{2}{c}{Gowalla-100k} & \multicolumn{2}{c}{Gowalla-1m} \\
\cmidrule(lr){2-3} \cmidrule(lr){4-5} \cmidrule(lr){6-7} \cmidrule(lr){8-9}
 & HR & NDCG & HR & NDCG & HR & NDCG & HR & NDCG \\
\midrule
\multicolumn{9}{c}{\textbf{Centralized Models}} \\
\midrule
MF & $0.0706_{\pm 0.0015}$ & $0.0286_{\pm 0.0012}$ & $0.1519_{\pm 0.0023}$ & $0.0957_{\pm 0.0018}$ & $0.1664_{\pm 0.0021}$ & $0.1461_{\pm 0.0019}$ & $0.1021_{\pm 0.0018}$ & $0.1085_{\pm 0.0016}$ \\
LightGCN & $0.0860_{\pm 0.0018}$ & $0.0339_{\pm 0.0014}$ & $0.1874_{\pm 0.0025}$ & $0.1177_{\pm 0.0020}$ & $0.1842_{\pm 0.0023}$ & $0.1658_{\pm 0.0021}$ & $0.1303_{\pm 0.0020}$ & $0.1435_{\pm 0.0018}$ \\
SGL & $0.0911_{\pm 0.0019}$ & $0.0349_{\pm 0.0015}$ & $0.1886_{\pm 0.0026}$ & $0.1188_{\pm 0.0021}$ & $0.1867_{\pm 0.0024}$ & $0.1692_{\pm 0.0022}$ & $0.1364_{\pm 0.0021}$ & $0.1471_{\pm 0.0019}$ \\
XSimGCL & $0.0959_{\pm 0.0020}$ & $0.0369_{\pm 0.0016}$ & $0.1767_{\pm 0.0024}$ & $0.1120_{\pm 0.0019}$ & $0.1896_{\pm 0.0025}$ & $0.1750_{\pm 0.0023}$ & $0.1466_{\pm 0.0022}$ & $0.1576_{\pm 0.0020}$ \\
\midrule
\multicolumn{9}{c}{\textbf{Federated Models}} \\
\midrule
FedMF & $0.0479_{\pm 0.0012}$ & $0.0176_{\pm 0.0008}$ & $0.0712_{\pm 0.0015}$ & $0.0443_{\pm 0.0012}$ & $0.1473_{\pm 0.0020}$ & $0.1193_{\pm 0.0018}$ & $0.0649_{\pm 0.0016}$ & $0.0685_{\pm 0.0014}$ \\
FedNCF & $0.0487_{\pm 0.0013}$ & $0.0202_{\pm 0.0009}$ & $0.0762_{\pm 0.0016}$ & $0.0465_{\pm 0.0013}$ & $0.1474_{\pm 0.0020}$ & $0.1296_{\pm 0.0019}$ & $0.0712_{\pm 0.0017}$ & $0.0822_{\pm 0.0015}$ \\
FedGNN & $0.0518_{\pm 0.0013}$ & $0.0202_{\pm 0.0009}$ & $0.1311_{\pm 0.0021}$ & $0.0789_{\pm 0.0017}$ & $0.1425_{\pm 0.0019}$ & $0.1266_{\pm 0.0018}$ & $0.0877_{\pm 0.0019}$ & $0.0968_{\pm 0.0017}$ \\
PerFedRec & $0.0549_{\pm 0.0014}$ & $0.0213_{\pm 0.0010}$ & $0.1338_{\pm 0.0022}$ & $0.0791_{\pm 0.0017}$ & $0.1365_{\pm 0.0018}$ & $0.1254_{\pm 0.0017}$ & $0.0878_{\pm 0.0019}$ & $0.0968_{\pm 0.0017}$ \\
PerFedRec++ & $0.0653_{\pm 0.0016}$ & $0.0246_{\pm 0.0011}$ & $0.1353_{\pm 0.0022}$ & $0.0784_{\pm 0.0017}$ & $0.1549_{\pm 0.0021}$ & $0.1388_{\pm 0.0019}$ & $0.0903_{\pm 0.0020}$ & $0.1023_{\pm 0.0018}$ \\
\textbf{FastPFRec} & $\mathbf{0.0705_{\pm 0.0017}^{*}}$ & $\mathbf{0.0271_{\pm 0.0012}^{*}}$ & $\mathbf{0.1399_{\pm 0.0023}^{*}}$ & $\mathbf{0.0861_{\pm 0.0018}^{*}}$ & $\mathbf{0.1696_{\pm 0.0022}^{*}}$ & $\mathbf{0.1494_{\pm 0.0020}^{*}}$ & $\mathbf{0.0997_{\pm 0.0019}^{*}}$ & $\mathbf{0.1073_{\pm 0.0017}^{*}}$ \\
\bottomrule
\end{tabular}
}
\label{tab:performance_merged_no_recall}
\end{table*}

To validate the effectiveness of our proposed method\footnote{Our source code is available at https://github.com/yanzhenxing123/FastPFRec}, we conducted a comparative study of FastPFRec against several competitive baselines, which include both centralized \cite{yu2023self} and federated recommendation methods. These baselines encompass traditional collaborative filtering techniques as well as GNN-based models. A detailed description of each baseline is provided below.

\noindent Centralized methods: 
\begin{itemize}[leftmargin=*]
\item
\textbf{MF \cite{koren2009matrix}:} It decomposes the user-item interaction matrix into low-dimensional user and item vectors, revealing the latent relationships between user preferences and item features.
\item
\textbf{LightGCN \cite{he2020lightgcn}: } It is a lightweight model that simplifies the aggregation of weight parameters in GCNs.
\item
\textbf{SGL \cite{wu2021self}:} It utilizes self-supervised learning to capture the relationships between users and items, improving recommendation accuracy by learning from the graph structure.
\item
\textbf{XSimGCL \cite{yu2023xsimgcl}:} It utilizes contrastive learning on cross-similarity graphs to enhance user and item representations. 
\end{itemize}
\noindent Federated recommendation methods:
\begin{itemize}[leftmargin=*]
\item
\textbf{FedMF \cite{chai2020secure}:}
It is a matrix factorization method based on FL, designed to address recommendation system problems in distributed scenarios.
\item

\textbf{FedNCF \cite{jiang2022fedncf}:}
It combines the strengths of matrix factorization and deep learning for improved recommendation accuracy in privacy-sensitive applications.
\item 

\textbf{FedGNN \cite{wu2021fedgnn}:}
It utilizes graph neural networks to facilitate feature interactions between users and items, enabling the model to effectively learn information from graph-structured data.
\item
\textbf{PerFedRec \cite{luo2022personalized}:} It combines joint representation learning, user clustering, and model adaptation to achieve improved recommendation performance and reduced communication costs. 
\item
\textbf{PerFedRec++ \cite{luo2024perfedrec++}:} PerFedRec++ is built upon the PerFedRec model, incorporated contrastive learning pre-training, resulting in superior performance for federated recommendation models.
\end{itemize}

\subsubsection{Implementation Details}\label{sec:Implementation_Details}

We evaluate FastPFRec and baseline methods in a simulated federated environment where multiple clients collaboratively train recommendation models without sharing raw user-item interaction data. To ensure a fair and reproducible comparison, all experiments are conducted on a workstation equipped with a NVIDIA Tesla V100 GPU, and 32GB RAM, running Ubuntu 20.04. The algorithms are implemented using Python 3.8 and PyTorch 2.2.2 with CUDA 11.8.

Regarding the experimental protocol, each experiment is repeated five times with different random seeds to ensure statistical reliability, and we report the results as the mean and standard deviation. We employ a consistent data splitting strategy, dividing the datasets into training, validation, and test sets with ratios of 70\%, 10\%, and 20\%, respectively. To ensure full reproducibility, we fix the random seed to 2025 for both data splitting and model initialization.

For hyperparameter settings, we utilize the federated recommendation framework with our proposed FastGNN as the backbone model. The parameters are configured as follows: the embedding dimension is set to $k=64$, the batch size to $B=256$, the learning rate to $\alpha=0.001$, and the regularization coefficient to $\gamma=0.0001$. The number of FastGNN layers is set to $H=2$. To balance training efficiency and performance, we configure the system with $T=10$ trusted nodes and use a ranking target of top-10. For baseline implementations, we integrate their official codebases into our federated framework and fine-tune their key hyperparameters on the validation set to ensure a fair comparison.

\begin{figure}[t]
\centerline{\includegraphics[width=0.45\textwidth]{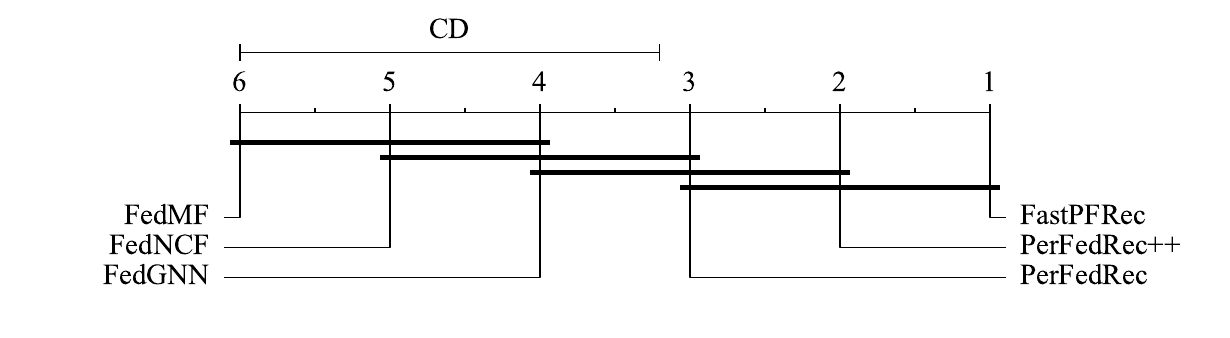}}
  %\vspace{-3mm}
\caption{CD diagram on HR.}
\label{fig:cd_graph}
  %\vspace{2mm}

\end{figure}

\subsection{Effectiveness of FastPFRec}\label{section_exp_effect}

{In this section, we evaluate the overall recommendation performance of FastPFRec against state-of-the-art baselines. The objective is to demonstrate that our framework achieves competitive or superior accuracy under standard federated settings before introducing strict privacy constraints.}

From the results in Table \ref{tab:performance_merged_no_recall}, it is evident that FastPFRec demonstrates significant performance advantages in federated recommendation tasks.

\begin{itemize}[leftmargin=*]
\item \textbf{Centralized Models:} Centralized models generally achieve better performance across all datasets due to their direct access to global data. For example, on the Gowalla-100k dataset, XSimGCL achieves an NDCG of 0.1576 and an HR of 0.1466, outperforming any federated models. However, centralized models depend heavily on full data sharing and storage, making them less applicable in scenarios that require privacy preservation.

\item \textbf{FastPFRec vs. Federated Baselines:} Among federated models, FastPFRec consistently outperforms other methods across all datasets and metrics. For instance, on the Gowalla-1m dataset, FastPFRec achieves higher metrics than PerFedRec and PerFedRec++. Federated GNN-based models perform better than FedMF, demonstrating the effectiveness of graph neural networks in capturing user-item relationships.

\item \textbf{Performance Improvements:} FastPFRec shows significant improvements over previous SOTA federated models. On the Gowalla-1m dataset, FastPFRec outperforms PerFedRec++ by 10.5\%, demonstrating strong adaptability in handling large-scale sparse data. These improvements highlight the effectiveness of FastPFRec's enhanced representation learning capabilities and privacy-preserving mechanisms.
\end{itemize}

Furthermore, we employed the Wilcoxon signed-rank test \cite{wilcoxon:Individual} to assess the significant differences between the algorithms and plotted the critical difference (CD) diagram, as shown in Figure \ref{fig:cd_graph}. The \textit{x}-axis represents the average ranking of each algorithm across multiple datasets based on the HR metric. If the difference in average rankings between the two algorithms exceeds the CD value, it indicates a statistically significant difference between them. From the plot, it is evident that FastPFRec is the best method with the lowest rank, significantly better than FedMF, FedNCF and FedGNN, while the current results are not sufficient to distinguish FastPFRec from PerFedRec and PerFedRec++.

The experimental results reveal varying performance improvements across datasets, which can be understood through their distinct characteristics (summarized in Table \ref{dataset}). FastPFRec achieves the most significant gains on the two Gowalla datasets. This is attributed to their larger scale and higher interaction density, which particularly benefit from the scheduled item-update mechanism. By updating a large number of item embeddings less frequently, substantial computational savings are realized without compromising accuracy, as the richer training signals enable embeddings to converge effectively even with sparse updates.

On the relatively smaller and sparser Yelp and Kindle datasets, the improvements are more modest yet consistent. Although the absolute efficiency gain is smaller due to the reduced problem scale, the method still maintains a superior efficiency-accuracy trade-off. The analysis of interaction patterns further explains these outcomes: datasets with more uniform distributions allow scheduled updates to work more harmoniously, whereas those with highly skewed popularity  still perform robustly, indicating that the update schedule generalizes well across different data distributions. Overall, the convergence speed and efficiency advantages of FastPFRec become increasingly pronounced as the dataset scale and density grow, demonstrating its strong suitability for large-scale recommendation scenarios.

\subsection{Ablation Study}\label{section_user}

{To understand the contribution of each algorithmic component, this section presents a detailed ablation study. We specifically examine how the FastGNN update strategy and the trusted-node aggregation mechanism independently affect the model's convergence and recommendation accuracy.}

As detailed in Table~\ref{tab:Ablation_study}, the results regarding structural optimizations offer critical insights. Removing the scheduled item updating strategy results in a significant performance drop, reducing accuracy to a level comparable to the standard LightGCN baseline. This confirms that without the scheduled update mechanism, the FastGNN architecture behaves similarly to a standard linearized GCN. Therefore, the superior performance of FastPFRec is primarily driven by the scheduled update strategy, which effectively mitigates gradient noise in non-IID federated settings.

Regarding security modules, the variants without trusted nodes and without anomaly detection demonstrate that our three-tier architecture is crucial for maintaining accuracy, as removing them leads to noticeable performance degradation. Furthermore, the variant without LDP noise represents the non-private baseline; as expected, it achieves the highest accuracy, validating the privacy-utility trade-off where our method sacrifices only 4.3\% utility for rigorous privacy. Finally, the update frequency analysis confirms that updating item embeddings every $H \times 10$ epochs strikes the optimal balance between model freshness and stability compared to frequencies of $H \times 5$ or $H \times 15$, and we set $H=2$.

\begin{table*}[ht]
\centering
\caption{Expanded Ablation Study: Performance of FastPFRec and its variants. "w/o" denotes the removal of a specific module. Note that "w/o LDP" represents the non-private baseline.}
\label{tab:Ablation_study}

\small 
\setlength{\tabcolsep}{4pt} 
\renewcommand{\arraystretch}{0.95} 

\begin{tabular}{lcccccc}
\toprule
\multirow{2}{*}{\textbf{Model / Variant}} & \multicolumn{2}{c}{\textbf{Yelp}} & \multicolumn{2}{c}{\textbf{Kindle}} & \multicolumn{2}{c}{\textbf{Gowalla-1m}} \\
\cmidrule(lr){2-3} \cmidrule(lr){4-5} \cmidrule(lr){6-7}
 & \textbf{HR@10} & \textbf{NDCG@10} & \textbf{HR@10} & \textbf{NDCG@10} & \textbf{HR@10} & \textbf{NDCG@10} \\
\midrule
\textbf{FastPFRec (Default)} & \textbf{0.0705} & \textbf{0.0271} & \textbf{0.1399} & \textbf{0.0861} & \textbf{0.0997} & \textbf{0.1073} \\
\midrule
\multicolumn{7}{l}{\textit{Structural Modules}} \\
\hspace{0.3cm} w/o FastGNN (use LightGCN) & 0.0642 & 0.0242 & 0.1331 & 0.0782 & 0.0912 & 0.0923 \\
\hspace{0.3cm} w/o Scheduled Item Updating (Full Update) & 0.0648 & 0.0246 & 0.1340 & 0.0790 & 0.0925 & 0.0938 \\
\midrule
\multicolumn{7}{l}{\textit{Security \& Privacy Modules}} \\
\hspace{0.3cm} w/o Trusted Nodes (Client-Server) & 0.0683 & 0.0261 & 0.1374 & 0.0851 & 0.0953 & 0.1033 \\
\hspace{0.3cm} w/o Anomaly Detection & 0.0691 & 0.0264 & 0.1382 & 0.0855 & 0.0975 & 0.1051 \\
\hspace{0.3cm} w/o LDP Noise ($\lambda=0$) & 0.0721 & 0.0285 & 0.1425 & 0.0885 & 0.1041 & 0.1120 \\
\midrule
\multicolumn{7}{l}{\textit{Update Frequency Analysis}} \\
\hspace{0.3cm} $H \times 5$  & 0.0689 & 0.0256 & 0.1382 & 0.0855 & 0.0987 & 0.1064 \\
\hspace{0.3cm} $H \times 15$ & 0.0695 & 0.0265 & 0.1385 & 0.0852 & 0.0982 & 0.1061 \\
\bottomrule
\end{tabular}
\end{table*}

\subsection{Convergence Analysis}\label{Convergence_proof}

In this section, we analyze the performance of FastPFRec with respect to the loss function and the number of epochs required for convergence analysis, with a particular focus on comparing its performance to baseline models.

\begin{figure}[!t]
\centering
\begin{minipage}[b]{0.49\columnwidth}
  \centering
  \includegraphics[width=\linewidth]{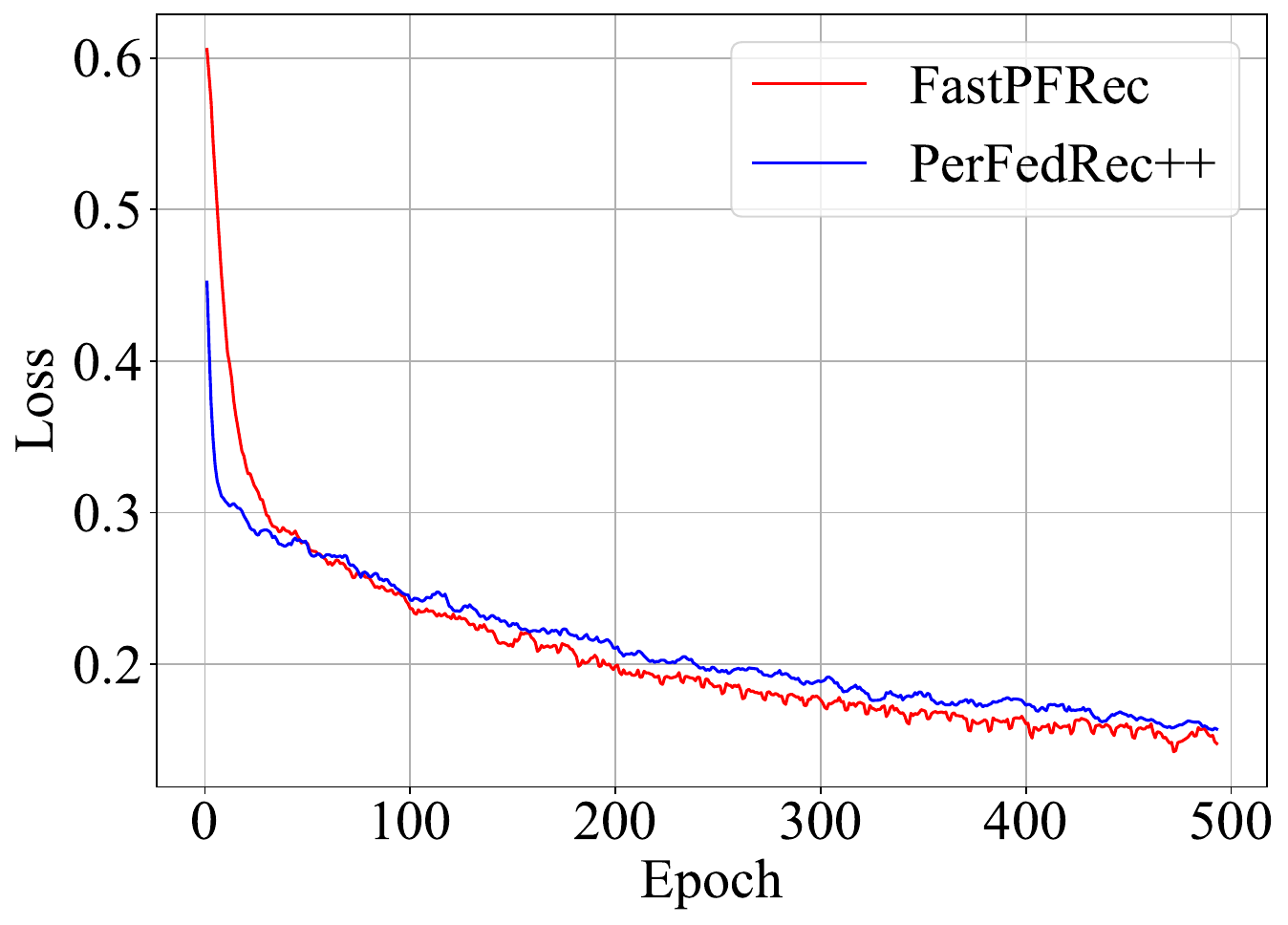}\\[-0.4em]
  \footnotesize (a) Training loss on Yelp.
  \label{subfig: loss_yelp}

\end{minipage}\hfill
\begin{minipage}[b]{0.49\columnwidth}
  \centering
  \includegraphics[width=\linewidth]{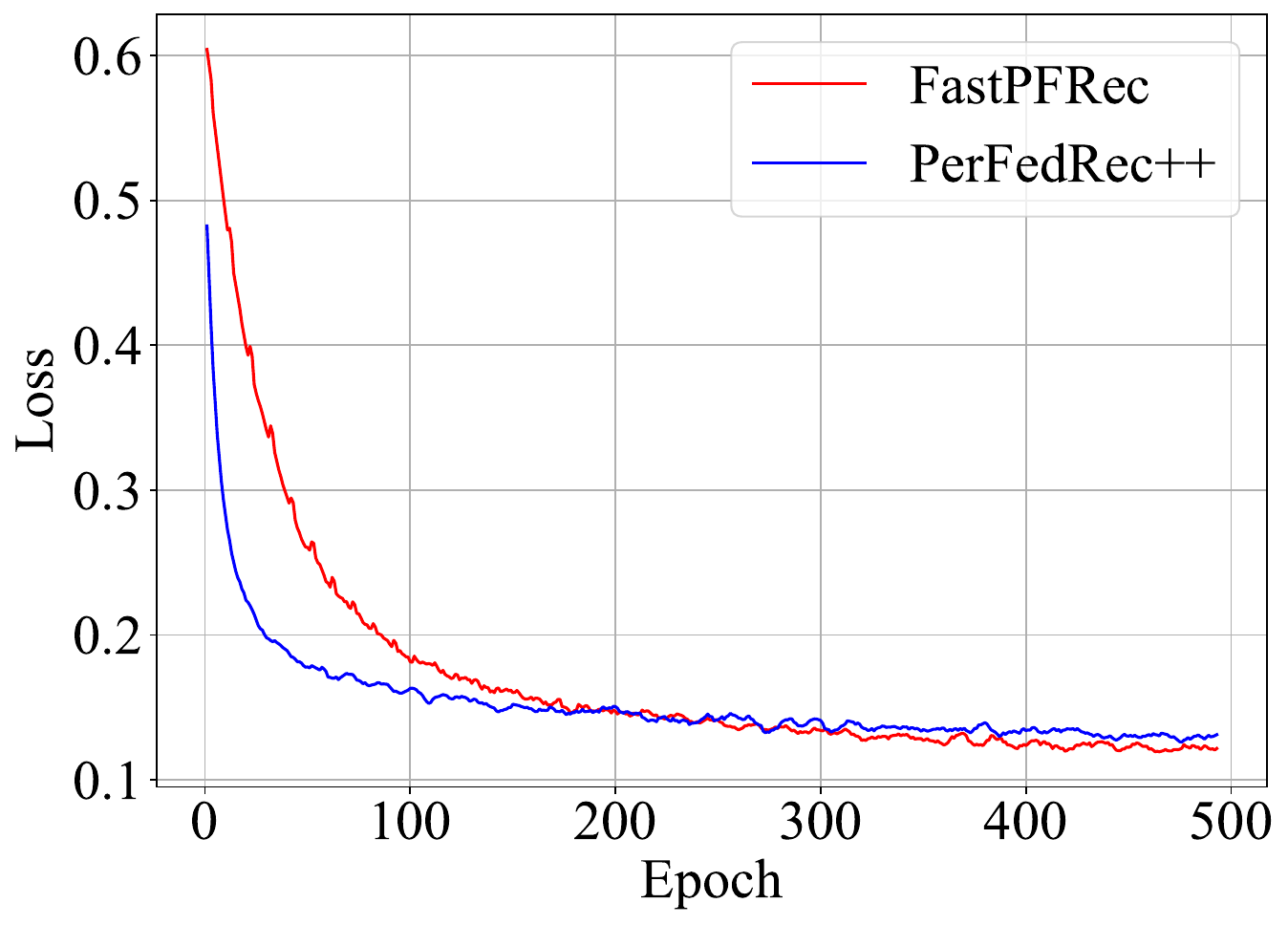}\\[-0.4em]
  \footnotesize (b) Training loss on Gowalla-1m.
    \label{subfig:loss_yelp_gowalla} % 主图像引用名称
\end{minipage}
\caption{Training loss of FastPFRec and PerFedRec++ on Yelp and Gowalla-1m.}
\label{fig:loss_yelp_gowalla}
\end{figure}

\begin{figure}[!t]
\centering
\begin{minipage}[b]{0.49\columnwidth}
  \centering
  \includegraphics[width=\linewidth]{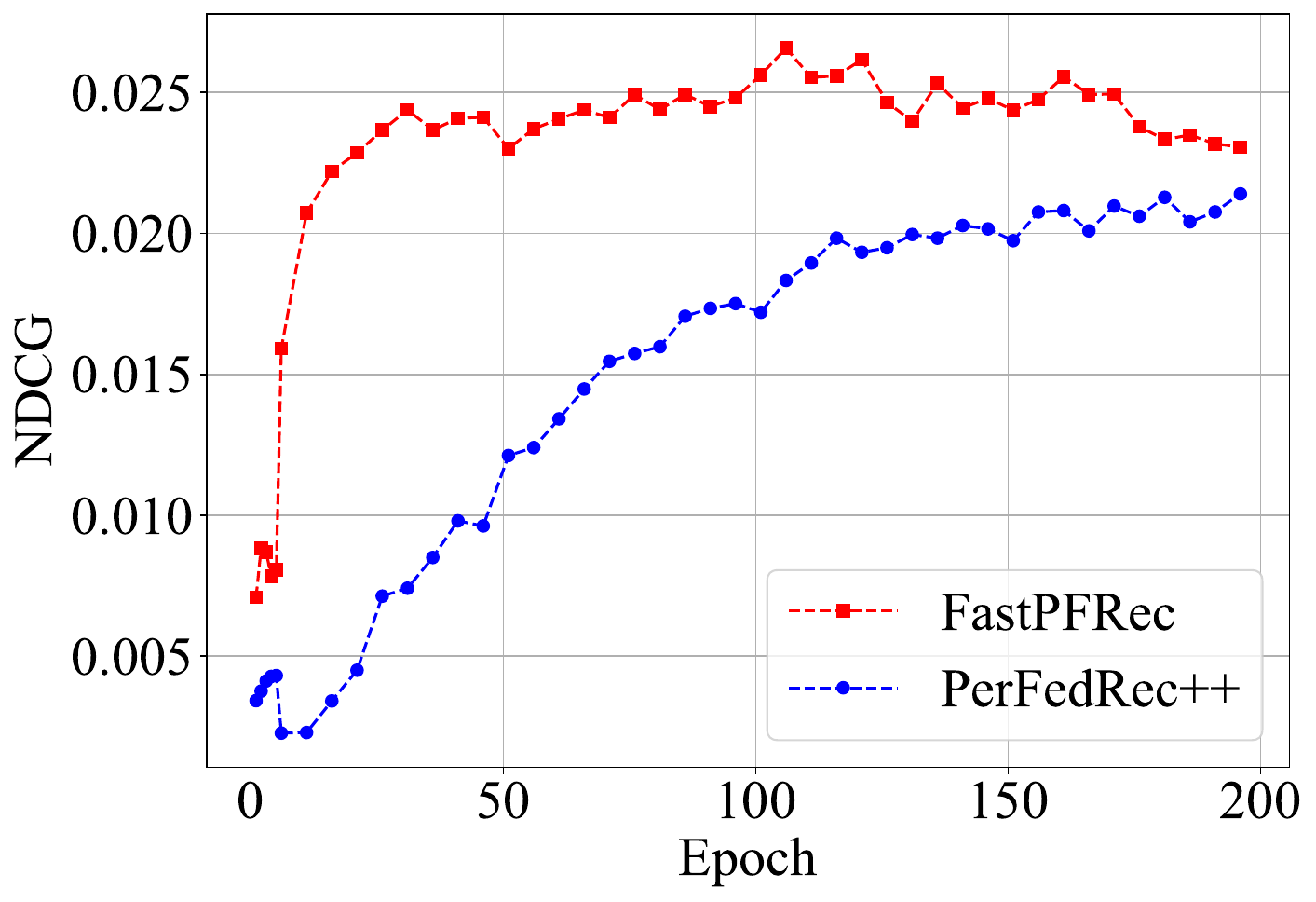}\\[-0.4em]
  \footnotesize (a) Training NDCG on Yelp.
 \label{subfig:ndcg_yelp}  % 子图像引用名称

\end{minipage}\hfill
\begin{minipage}[b]{0.49\columnwidth}
  \centering
  \includegraphics[width=\linewidth]{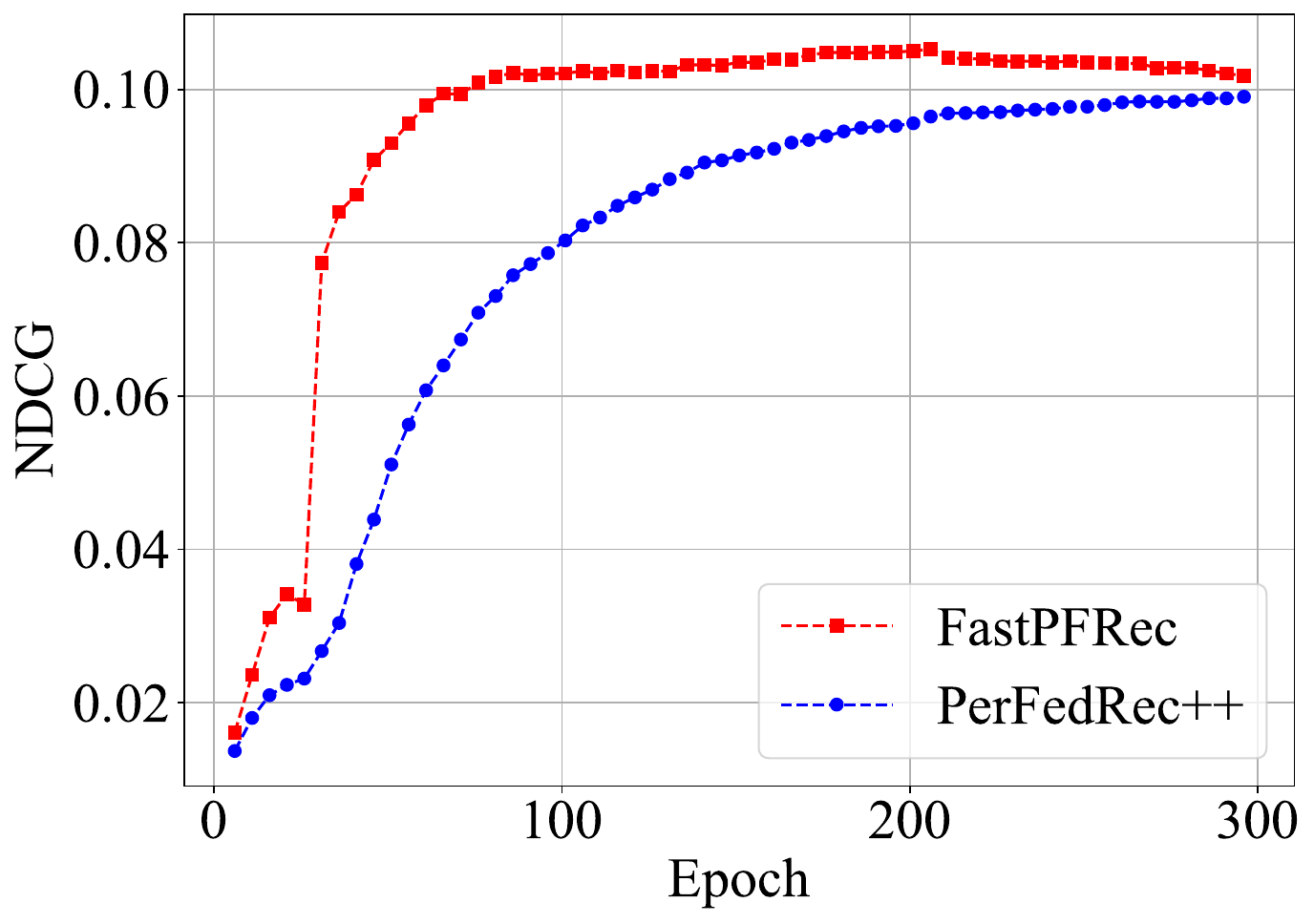}\\[-0.4em]
  \footnotesize (b) Training NDCG on Gowalla-1m.
    \label{fig:epoch_ndcg_yelp} % 主图像引用名称

\end{minipage}
\caption{Training NDCG of FastPFRec and PerFedRec++ on Yelp and Gowalla-1m (per epoch).}
\label{fig:epoch_ndcg}
\end{figure}

In Figure~\ref{fig:loss_yelp_gowalla}, we compare the loss curves of FastPFRec and PerFedRec++ across different datasets for each training epoch. Our experimental results show that FastPFRec consistently achieves lower loss values across all evaluation metrics than the baseline model. Specifically,  FastPFRec exhibits a smoother convergence process and a more stable learning curve, whereas the baseline model tend to fluctuate more significantly. This indicates that our model is not only more efficient in terms of performance but also demonstrates better generalization to unseen data. 

\begin{figure}[t]
\centerline{\includegraphics[width=0.45\textwidth]{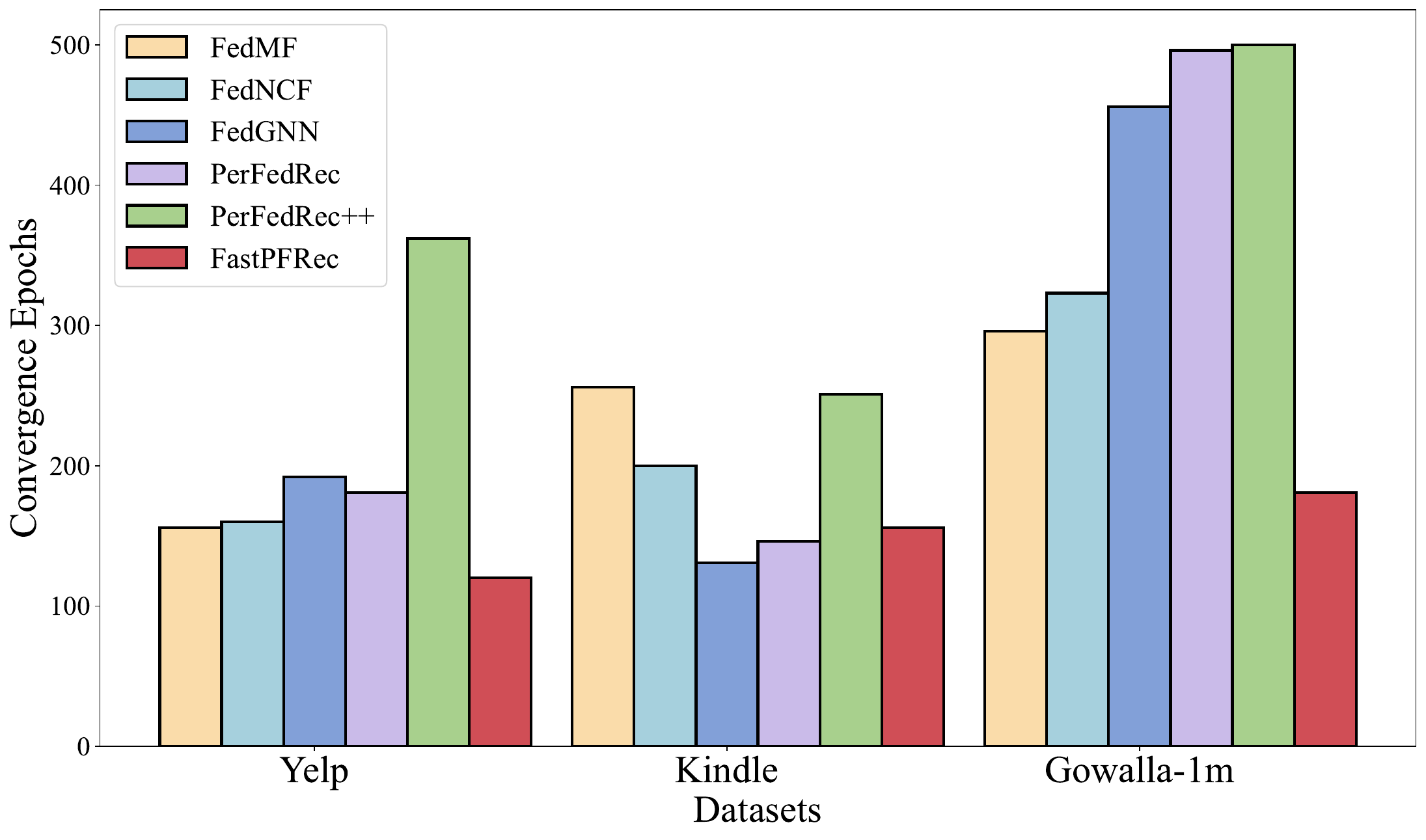}}
  % \vspace{-3mm}
  \caption{Comparison of epochs for federated recommendation models achieving the best NDCG across different datasets.}
  \label{fig:ZZT_ndcg_all}
  % \vspace{-6mm}
\end{figure}

Furthermore, we analyze the number of epochs required for the model to reach its best performance. In Figure~\ref{fig:epoch_ndcg}, we compare the changes in the NDCG score for FastPFRec and PerFedRec++ across different datasets. Additionally, Figure~\ref{fig:ZZT_ndcg_all} compares the number of epochs required for FastPFRec and other federated recommendation baselines to achieve their best performance across various datasets.

As shown in Figure \ref{fig:epoch_ndcg} (a), our experiments demonstrate that FastPFRec converges significantly faster than the baseline model, achieving approximately optimal performance in just 50 epochs on the Yelp dataset. This indicates the model’s efficiency in learning and adapting to the data. In contrast, the baseline model PerFedRec++ typically requires more epochs to reach a comparable level of performance, i.e., taking up to 200 epochs to achieve similar results. The faster convergence of FastPFRec can be attributed to its improved GNN interaction mechanism, which helps it more effectively learn the underlying patterns in the data.

On larger and more complex datasets, FastPFRec maintains its superior convergence characteristics. As shown in Figure \ref{fig:epoch_ndcg} (b), on large-scale datasets, FastPFRec achieves its best performance around 80 epochs, whereas baseline models typically require 300 or even 500 epochs to reach optimal results. Similar results are indicated in the comparison with other baselines in Figure \ref{fig:ZZT_ndcg_all}. Compared to the baselines, we achieve the best performance with 32.0\% fewer training epochs. This accelerated convergence is especially valuable in real-world applications where training time is a critical factor. The reduced training time allows FastPFRec to be deployed more quickly and efficiently, making it a highly practical solution for large-scale recommendation tasks.

{As evidenced by these convergence curves, the scheduled item-update mechanism effectively prevents the rapid propagation of localized gradient noise, leading to the remarkably stable and accelerated convergence of FastPFRec. A deeper theoretical discussion regarding how this mechanism prevents over-smoothing is provided in Section 4.10.}

\subsection{Privacy-Utility Trade-off Analysis}
\label{section_tradeoff}

To explicitly address the concern regarding the impact of privacy mechanisms on recommendation quality, we conducted a sensitivity analysis on the Gowalla-1m dataset. We compared our fully private model against a baseline where LDP noise was disabled, quantifying the specific cost of differential privacy.

Table~\ref{tab:privacy_tradeoff} presents the detailed impact of varying the privacy parameters. 
The baseline configuration (with graph perturbation $P_\text{pert}=0.1$ but no LDP noise) achieves an NDCG of {0.1120}. 
When we enforce our default LDP noise ($\lambda=0.1$), the performance decreases to {0.1073}. 
This represents a total utility cost of approximately {4.29\%}. 

It is worth noting that further increasing the noise scale to $\lambda=0.2$ only leads to a marginal additional drop (to 0.1065), demonstrating that the FastPFRec architecture is relatively robust to parameter perturbation. 
Even with these strict privacy constraints, our method (0.1073) continues to outperform the state-of-the-art federated baseline PerFedRec++ (0.1023), confirming that the proposed trade-off provides essential security guarantees without compromising service quality.

\begin{table}[h]
\centering
\caption{Privacy-Utility Trade-off on Gowalla-1m. We fix the graph perturbation $P_\text{pert}=0.1$ and vary the LDP noise scale $\lambda$. The row with $\lambda=0.10$ corresponds to our default setting.}
\label{tab:privacy_tradeoff}

\small 
\setlength{\tabcolsep}{6pt}
\renewcommand{\arraystretch}{0.95} 
% --------------------------

\begin{tabular}{cc|cc}
\toprule
\multicolumn{2}{c|}{\textbf{Privacy Parameters}} & \multicolumn{2}{c}{\textbf{Utility Metric}} \\
$P_\text{pert}$ & $\lambda$ (LDP Noise) & NDCG & Drop Rate (\%) \\
\midrule
0.10 & 0 (No Noise) & \textbf{0.1120} & - \\
\midrule
0.10 & 0.05 & 0.1108 & 1.07\% \\
\textbf{0.10} & \textbf{0.10 (Default)} & \textbf{0.1073} & \textbf{4.29\%} \\
0.10 & 0.20 & 0.1065 & 4.91\% \\
\bottomrule
\end{tabular}
\end{table}

\subsection{Trusted Node Security and Resilience Analysis}\label{subsec:attack_resilience}

This section comprehensively evaluates the security and resilience of the trusted node architecture, examining its effectiveness against malicious attacks and its robustness under infrastructure failures.

We first validate the effectiveness of our trusted node architecture in mitigating malicious attacks through a controlled experiment. We simulate a scenario where 30\% of clients are malicious and inject Gaussian noise into their local model parameters. We compare two system architectures: a traditional client-server model and our three-tier architecture with trusted nodes. The experiment uses 200 clients distributed across 10 trusted nodes, with 30 independent trials.

As shown in Table~\ref{tab:attack_resilience}, the trusted node architecture achieves a 94.8\% reduction in server damage compared to the traditional client-server model, where server damage is quantified by the L2-norm $\|\mathbf{W}_{new} - \mathbf{W}_{old}\|_2$. The architecture successfully detects and isolates 94.8\% of malicious behaviors, preventing attack propagation to the central server and ensuring system stability even under high malicious client ratios.

% \textbf{Infrastructure Resilience Evaluation.}

Beyond deliberate attacks, we evaluate system resilience under infrastructure failures using the Gowalla-100K dataset. Table \ref{tab:node_failure} shows system performance under different numbers of failed trusted nodes. With $T=10$ trusted nodes, the system maintains robust performance even with 30\% node failure, exhibiting only minor accuracy degradation (less than 2.0\%). This confirms the architecture's inherent redundancy and ability to gracefully degrade under infrastructure faults.

Table \ref{tab:node_compromise} further presents attack resilience when trusted nodes themselves are compromised to launch noise-injection attacks. The anomaly detection system effectively contains such threats, maintaining model accuracy within 1.5\% of the baseline and achieving containment rates 98.7\%, even when 30.0\% of nodes are malicious. Table \ref{tab:detection_performance} details the effectiveness of our MAD-based detection, which achieves an average detection rate exceeding 96.0\% with false positive rates below 5.0\%, while introducing minimal latency overhead (under 50 ms).

% \textbf{Security Design Validation.}
Collectively, these results validate the multi-faceted security design of FastPFRec. The three-tier architecture, combined with proactive anomaly detection and redundant infrastructure, provides strong resilience against both deliberate malicious attacks and accidental infrastructure failures. This ensures reliable and secure operation in practical federated learning deployments, making the architecture particularly suitable for privacy-sensitive applications requiring robust security guarantees.

\begin{table}[ht]
\centering
\small 
\caption{Attack resilience comparison: Traditional vs. FastPFRec architecture}
\label{tab:attack_resilience}

\setlength{\tabcolsep}{3pt} 
% --------------------------

\begin{tabular}{ccc}
\toprule
{Architecture} & {Server Damage Avg} & {Protection Rate} \\
\midrule
Traditional (Direct) & $2.34 _{\pm 0.15}$ & 0.0\% \\
FastPFRec (w/ Trusted Nodes) & $0.12 _{\pm 0.08}$ & 94.8\% \\
\bottomrule
\end{tabular}
\end{table}

\begin{table}[ht]
\centering
\small
\caption{System performance under trusted node failures}
\label{tab:node_failure}
\begin{tabular}{ccc}
\toprule
{Failed Nodes} & {Failure Rate} & {NDCG} \\
\midrule
0 & 0\% & $0.1073 _{\pm 0.0002}$ \\
1 & 10\% & $0.1068_{ \pm 0.0004}$ \\
2 & 20\% & $0.1070 _{\pm 0.0008}$ \\
3 & 30\% & $0.1064 _{\pm 0.0005}$ \\
\bottomrule
\end{tabular}
\end{table}

\begin{table}[ht]
\centering
\small
\caption{Attack resilience under node compromise}
\label{tab:node_compromise}
\begin{tabular}{ccc}
\toprule
{Compromised Nodes} & {NDCG} & {Containment Rate} \\
\midrule
0 & $0.1073 _{\pm 0.0013}$ & 100.0\% \\
1 & $0.1069 _{\pm 0.0009}$ & 99.2\% \\
2 & $0.1065 _{\pm 0.0010}$ & 98.7\% \\
3 & $0.1062_{ \pm 0.0008}$ & 98.1\% \\
\bottomrule
\end{tabular}
\end{table}

\begin{table}[ht]
\centering
\small
\caption{Anomaly detection performance metrics}
\label{tab:detection_performance}

\setlength{\tabcolsep}{4pt} 

\begin{tabular}{cccc}
\toprule
{Attack Type} & {Detection Rate} & {False Positive} & {Latency (ms)} \\
\midrule
Noise Injection & 97.3\% & 3.2\% & $38 \pm 5$ \\
Gradient Poisoning & 96.8\% & 4.1\% & $46 \pm 6$ \\
\bottomrule
\end{tabular}
\end{table}

\subsection{Efficiency Analysis}\label{section:Efficiency_Analysis}

{This section evaluates the computational efficiency and training acceleration introduced by the FastGNN update strategy. The primary objective is to empirically quantify the reduction in training time and computational overhead compared to conventional full-graph update methods, thereby demonstrating the practical scalability of our approach.}

\begin{figure}[t]
    \centerline{\includegraphics[width=0.45\textwidth]{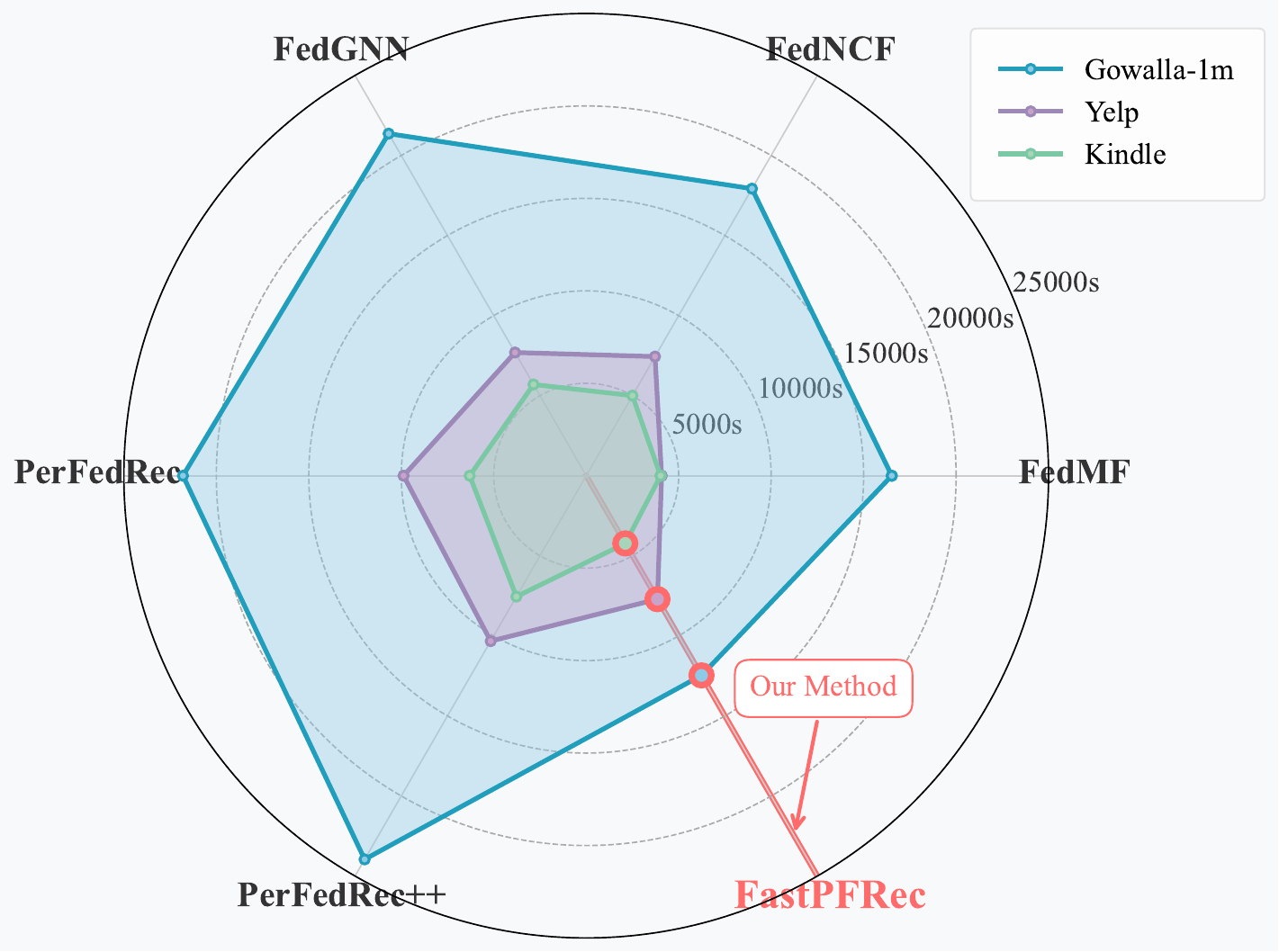}}
    % \vspace{-2mm}
    \caption{Time consumption of federated recommendation models over same epochs.}
    \label{fig:Model_Execution_Time}
    % \vspace{-2mm}
\end{figure}

Specifically, to analyze the time efficiency of FastPFRec, we compare its runtime and time complexity with state-of-the-art methods. Figure~\ref{fig:Model_Execution_Time} reports the per-epoch runtime across datasets. FastPFRec achieves the lowest time consumption, including a 43.0\% reduction on Gowalla-1m compared to PerFedRec++. Averaged over all datasets, it reduces runtime by 34.1\% relative to GNN-based federated baselines.

\begin{table}[ht]
\centering
\small
\vspace{-4mm}
\caption{Time complexity of different federated client models (per epoch).}
\begin{tabular}{cc}
\toprule
{Model} & Time Complexity \\
\midrule
FedMF & $\mathcal{O}(|U| \cdot |I| \cdot k)$ \\
FedNCF & $\mathcal{O}(H \cdot |U| \cdot |I| \cdot k)$ \\
FedGNN & $\mathcal{O}(H \cdot |U| \cdot |I| \cdot k)$  \\ 
PerFedRec  & $\mathcal{O}(H \cdot |U| \cdot |I| \cdot k)$\\ 
PerFedRec++ & $\mathcal{O}(H \cdot |U| \cdot |I| \cdot k)$ \\
FastPFRec & $\mathcal{O}(H \cdot k \cdot (|U| + |I|/h))$ \\
\bottomrule
\end{tabular}
\label{table:Time_complexity}
\end{table}

Table~\ref{table:Time_complexity} shows the total computational cost of the learning representation in the federated recommendation, where $H$ is the number of convolutional layers, $k$ is the vector dimension, $|U|$ and $|I|$ denote user and item counts. For the GNN-based methods (i.e., FedGNN, PerFedRec, PerFedRec++, and FastPFRec), we only consider the time complexity of their local GNN modules. 

In our model, when performing FastGNN convolution, the user-item interaction matrix is utilized as the local graph. During the convolution process, the user matrix convolves with the item matrix at every step, while feature interaction from items to users occurs only every $H\times10$ epochs. This scheduled strategy reduces the amortized computational cost of item embedding updates by a factor of approximately $H \times 10$ compared to full updates, resulting in a linear reduction in overall complexity as shown in Table~\ref{table:Time_complexity}.

\subsection{Hyperparameter Analysis}\label{Hyperparameter_Analysis}

\begin{figure}[!t]
\centering
\begin{minipage}[b]{0.49\columnwidth}
  \centering
  \includegraphics[width=\linewidth]{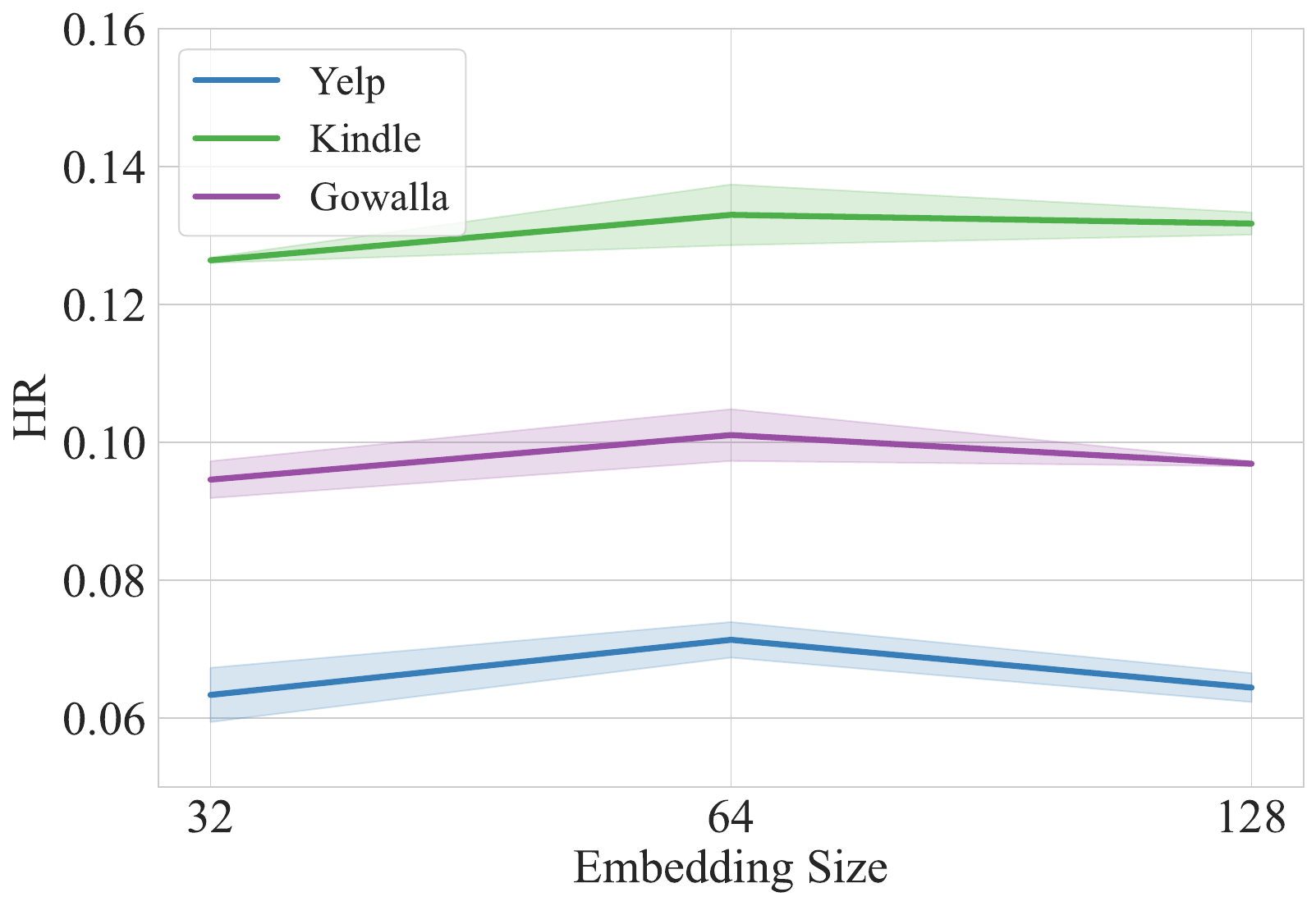}\\[-0.4em]
  \footnotesize (a) HR on different datasets.
  \label{subfig:hr_emb_size}  % 子图像引用名称

\end{minipage}\hfill
\begin{minipage}[b]{0.49\columnwidth}
  \centering
  \includegraphics[width=\linewidth]{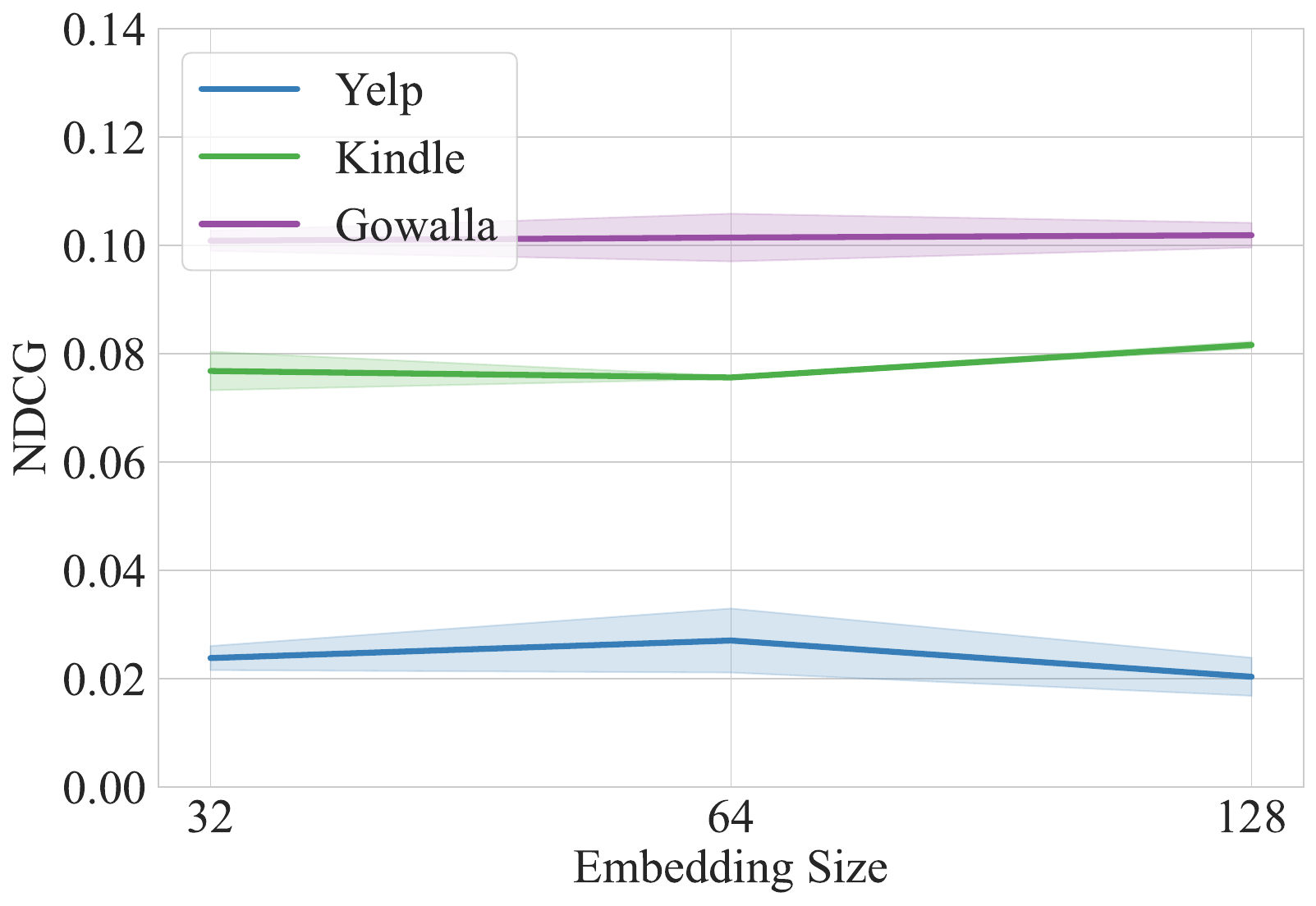}\\[-0.4em]
  \footnotesize (b) NDCG on different datasets.
  \label{subfig:ndcg_recall_emb_size} % 主图像引用名称

\end{minipage}
\caption{Performance comparison with different embedding size.}
\label{fig:ndcg_recall_emb_size}
\end{figure}

This section analyzes the impact of hyperparameters, with a particular focus on the embedding size $k$ and the number of trusted nodes $T$.

To analyze the effect of embedding size on model performance, we conducted experiments on the Yelp, Kindle, and Gowalla-1m datasets, using three different embedding size configurations: 32, 64, and 128. The experimental results for HR and NDCG at different embedding sizes are shown in Figure~\ref{fig:ndcg_recall_emb_size}. As the embedding size increases from 32 to 64, all datasets show improvements in both HR and NDCG, with the Kindle dataset achieving the highest peak performance. However, further increasing the embedding size to 128 results in performance degradation for most datasets, likely due to overfitting or feature redundancy. Gowalla-1m exhibits the most robust performance, with relatively stable NDCG values. These results highlight the importance of selecting an optimal embedding size, typically around 64, to balance performance and model complexity.

Table~\ref{trusted_nodes_nums} shows the relationship between the number of trusted nodes and performance and time cost (indicated by $\mathcal{T}_{500}$) under training 500 epochs. As the number of trusted nodes increases from 5 to 10, there is a slight improvement in performance for all datasets. However, further increasing the number to 20 shows marginal gains or even minor declines in certain cases, indicating diminishing returns. For training time, as the number of clients increases, the training time will increase slightly. This suggests that while increasing trusted nodes can enhance performance, the improvements are limited and may come at the cost of reduced training efficiency. Hence, a balance must be struck between performance improvement and computational efficiency.

\begin{table*}[ht]
\centering  % This centers the table
\small
\caption{Performance and time cost (in seconds) comparison with different numbers of trusted nodes.}
\label{trusted_nodes_nums}

\begin{tabular}{cccccccccc}
\toprule
\multirow{2}{*}{$T$} & \multicolumn{3}{c}{Yelp} & \multicolumn{3}{c}{Kindle} & \multicolumn{3}{c}{Gowalla-1m} \\
\cmidrule(lr){2-4} \cmidrule(lr){5-7} \cmidrule(lr){8-10}
 & HR & NDCG & $\mathcal{T}_{500}$ & HR & NDCG & $\mathcal{T}_{500}$ & HR & NDCG & $\mathcal{T}_{500}$ \\
\midrule
{5} & {0.0688} & {0.0261} & {3303} & {0.1374} & {0.0851}  &{3194} & {0.0973} & {0.1032} & 21933 \\
{10}  & {0.0704} & {0.0270} &  {3429} &{0.1398} & {0.0860} & {3204} & {0.0996} & {0.1073} & {22563}  \\
{20} & {0.0682} & {0.0268}  & {3518} & {0.1361} & {0.0858}  & {3317} & {0.0972} & {0.1029}  & {24016} \\

\bottomrule
\end{tabular}
% }
\end{table*}

\subsection{Real-World Federated Simulation}
In this section, we construct a simulated real-world federated recommendation environment to fully evaluate the communication efficiency of FastPFRec. The purpose of this simulation is to emulate practical FL scenarios, in which the cost of data transmission between clients, intermediate trusted nodes, and a central server is often a critical bottleneck. Inspired by HN3S \cite{ZHANG2024103580}, the details of simulation are described below.

We train all models for 500 epochs with a batch size of 256 on a system running Ubuntu 20.04, equipped with an NVIDIA Tesla V100 GPU and a 1TB SSD. During training, we explicitly record only the time spent on communication-related operations, including:
\begin{itemize}[leftmargin=*, itemsep=0pt, topsep=0pt]
    \item Transmission of model parameters between clients, trusted nodes, and the server.
    \item Aggregation time at trusted nodes and the central server.
    \item Loading and dispatching time of the model updates.
\end{itemize}
As shown in Table~\ref{tab:comm_time}, we report the communication costs explicitly. Note that computation time (e.g., forward/backward propagation) is excluded to isolate communication overhead.

\begin{table}[ht]
\centering
\small
\caption{Communication time in real-world simulation (in seconds)}
\begin{tabular}{cccc}
\toprule
Model & Yelp & Kindle & Gowalla-1m \\
\midrule
FastPFRec & 9.35 & 8.23 & 12.09 \\
PerFedRec++ & 9.59 & 7.87 & 13.24 \\
\bottomrule
\end{tabular}
\label{tab:comm_time}
\end{table}
These results confirm that the integration of trusted nodes does not introduce substantial communication overhead. On the contrary, the decentralized aggregation mechanism and lightweight update strategy of FastPFRec contribute to {efficient communication}, making the method suitable for large-scale, real-world federated recommendation systems where low latency and scalability are critical.

\subsection{Discussion}
The superior convergence and accuracy of FastPFRec compared to existing GNN-based federated methods (e.g., FedGNN and PerFedRec++) are fundamentally driven by its scheduled item-update mechanism. In collaborative bipartite graphs, frequent full-graph convolutions inevitably lead to the well-known over-smoothing problem \cite{chen2020measuring, li2018deeper}, where item embeddings lose their distinctiveness by repeatedly absorbing heterogeneous user information. Since item characteristics are intrinsically more stable than dynamic user preferences, FastGNN strategically reduces the frequency of item convolutions. This asynchronous update mechanism effectively prevents over-smoothing, allowing user embeddings to adapt dynamically while preserving the unique, stable representational capacity of items. Consequently, the global model converges faster and avoids the performance degradation typical of over-convolved GNNs.

Furthermore, our three-tier architecture offers systemic security advantages over standard two-tier systems. By utilizing trusted nodes that are deployed as reliable, infrastructure-level edge servers, FastPFRec logically and physically decouples clients from the central server. This architecture effectively neutralizes server-side source-tracing vulnerabilities and isolates malicious updates. Crucially, this design exhibits strong fault tolerance; empirical results demonstrate that the system maintains robust performance even under a $30\%$ node failure or compromise scenario, as the adverse impact remains strictly contained within isolated subsets of clients.

Regarding the privacy-utility trade-off, empirical results indicate a marginal performance degradation (e.g., a $4.29\%$ NDCG drop on the Gowalla-1m dataset) when enforcing strict LDP constraints. In practical deployments, this trade-off is highly acceptable and often a fundamental prerequisite for compliance with stringent data protection regulations, such as the GDPR. Furthermore, it is noteworthy that even under these strict privacy constraints ($\lambda=0.1$), FastPFRec consistently outperforms state-of-the-art baselines (e.g., PerFedRec++) that operate entirely without privacy noise. This demonstrates that our architectural design effectively bounds the utility cost of privacy preservation, offering a superior balance compared to alternative approaches that often suffer severe accuracy drops when perturbation-based privacy is applied.

Despite these advantages, the framework has certain limitations. Its security guarantees inherently rely on the ``honest-but-curious'' assumption \cite{yang2019federated} of these infrastructure-level trusted nodes, making deployment challenging in fully decentralized ad-hoc environments where such reliable intermediaries are unavailable. Additionally, while the system is highly robust against noise-injection attacks, defending against more sophisticated adversarial threats, such as advanced model inversion or membership inference attacks, remains an open challenge for future research.

\section{Conclusion}
This study presents FastPFRec, a fast and personalized federated recommendation framework with integrated privacy protection designed to address data privacy and heterogeneity challenges. The experimental results across multiple datasets demonstrate that FastPFRec achieves high recommendation accuracy while significantly improving training efficiency through the FastGNN update schedule and providing robust privacy guarantees through a multi-layered protection mechanism. The three-tier architecture with trusted nodes further enhances security against malicious clients and shows strong scalability, particularly on large-scale datasets.

Future work will focus on: (1) exploring efficient model compression techniques and further algorithmic optimizations to enhance scalability with large numbers of users and items; (2) developing more advanced differential privacy strategies and evaluating the framework under a broader range of privacy attacks; and (3) improving the trusted-node aggregation mechanism and extending the framework to scenarios without trusted intermediate infrastructure.

\section*{Acknowledgments}
This work is supported in part by the Fundamental Research Funds for the Central Universities (No. 2025JBMC028), and in part by the National Natural Science Foundation of China (No. 61702030).

\bibliographystyle{elsarticle-num-names} 
\bibliography{egbib1}

\end{document}